\documentclass[sigconf]{acmart_changed}
\usepackage{amsmath}
\usepackage[noend]{algorithmic}
\usepackage[nameinlink,capitalize]{cleveref}
\usepackage{float}
\usepackage{algorithm}
\usepackage{subcaption}
\usepackage{dsfont}
\usepackage{multirow}
\usepackage{array}
\usepackage{enumitem}
\usepackage[resetlabels]{multibib}

\newcites{AP}{Appendix References}
\usepackage{balance}
\usepackage{booktabs}
\usepackage{rotating}
\usepackage{tikz}
\usepackage{soul}
\definecolor{navy}{rgb}{0.1, 0.1, 0.8}
\definecolor[named]{gray}{rgb}{0.4, 0.4, 0.4}
\definecolor[named]{olive}{rgb}{0.1, 0.5, 0.1}
\definecolor[named]{ruby}{rgb}{0.8, 0.1, 0.3}
\definecolor{darkpastelgreen}{rgb}{0.01, 0.75, 0.24}
\definecolor{celestialblue}{rgb}{0.29, 0.59, 0.82}
\definecolor{coral}{rgb}{1.0, 0.5, 0.31}
\definecolor{Goldenrod}{rgb}{0.8,0.8,0}

\DeclareMathOperator{\Prob}{\mathbb{P}}

 \newcommand{\eat}[1]{}

\setcopyright{none}
\renewcommand\footnotetextcopyrightpermission[1]{} %
\begin{document}
\title{Contact Tracing: Computational Bounds, Limitations and Implications}

\author{Quyu Kong}
\authornote{Work done during an internship at the Max Planck Institute for Human Development}
\affiliation{%
  \institution{Australian National University \&\\ Data61, CSIRO \& UTS}
}
\email{quyu.kong@anu.edu.au}

\author{Manuel Garcia-Herranz}
\affiliation{%
  \institution{UNICEF}
  }
\email{mgarciaherranz@unicef.org}

\author{Ivan Dotu}
\authornote{Corresponding authors}
\affiliation{%
  \institution{Moirai Biodesign}
  }
\email{ivan.dotu@moiraibiodesign.com}

\author{Manuel Cebrian}
\authornotemark[2]
\affiliation{%
  \institution{Max Planck Institute for Human Development}
  }
\email{cebrian@mpib-berlin.mpg.de}

\begin{abstract}
Contact tracing has been extensively studied from different perspectives in recent years. However, there is no clear indication of why this intervention has proven effective in some epidemics (SARS) and mostly ineffective in some others (COVID-19). Here, we perform an exhaustive evaluation of random testing and contact tracing on novel superspreading random networks to try to identify which epidemics are more containable with such measures. We also explore the suitability of positive rates as a proxy of the actual infection statuses of the population. Moreover, we propose novel ideal strategies to explore the potential limits of both testing and tracing strategies. Our study counsels caution, both at assuming epidemic containment and at inferring the actual epidemic progress, with current testing or tracing strategies. However, it also brings a ray of light for the future, with the promise of the potential of novel testing strategies that can achieve great effectiveness.
\end{abstract}

\fancyhead{}
\maketitle
\section*{Introduction}\label{sec:questions}

The COVID-19 pandemic has posed significant threats to public health globally since 2020. It spreads rapidly around the world and it is hard to control due to a large proportion of pre-symptomatic and asymptomatic cases~\citep{mizumoto2020estimating}. Although preventive measures, such as social distancing~\citep{Prem2020} and lockdowns~\citep{Qian2020,flaxman2020estimating}, have shown to be effective in slowing down the disease spreading, they come at the cost of economic downturn and the risk of a second wave after lifting restrictions. In order to respond to such time-critical emergencies while avoiding strict measures, testing of potential infections via contact data and quarantining of positive cases have been proposed as alternative measures~\citep{Aleta2020,Lorch2020}. However, limited testing/tracing resources have imposed challenges on the mass deployment of such measures.

In this work, we tackle an extensive evaluation of different random testing and contact tracing strategies to ascertain their efficiency. To answer this, we first build a simulation tool involving two major components --- an extended stochastic epidemic model simulator with tracing and quarantining actions, and a contact network simulator that takes both $R_0$ and the dispersion parameter into account. These allow us to identify the characteristics of each strategy for a large set of parameter combinations. Moreover, by introducing two ideal (despite unreal) strategies, we can propose a classification of an epidemic (along with the initial time point and testing resources) with levels: 
1) can be contained with classic contact tracing (either forward, backward or both), 2) can be contained with contact tracing with priorities, 3) can be contained with smart testing strategies and 4) cannot be contained with testing.

Overall, we find that: our superspreading networks better characterize the actual dispersion effect; backward contact tracing is slightly better at finding positive cases for epidemics with small dispersion parameter with a small number of tests; and that there is both a gap between classic contact tracing and ideal contact tracing and between it and ideal testing, that make exploring smart contact tracing and testing strategies worth it. We also show the limitations of contact tracing as a proxy for the actual epidemic status and propose random testing as a better estimate for such inference. Finally, our models assume full knowledge of the contact network, akin to using a perfect digital tracking app; other modeling decisions attempt at conveying optimal computational conditions for contact tracing and testing.

\subsection*{Background}\label{sec:models}
Compartmental epidemic models, such as the Susceptible-Infected-Recovered (SIR) model~\citep{kermack1927contribution}, are classical models used for gaining insights into disease transmission dynamics. Such models define several infection states for individuals in a population, and the dynamics of individuals moving between states can be modeled deterministically or stochastically~\citep{allen2008mathematical}.

Many works have explored epidemic simulations over random networks~\citep{newman2002spread,kiss2017mathematics,newman2018networks}. The key issues to consider in this context are the stochasticity of epidemic diffusions and the network architecture.

Stochastic simulations of epidemics are conducted with the Gillespie algorithm~\citep{kiss2017mathematics}. The algorithm samples, at each step, from the total rates of all possible events, the next event time and an event type---e.g., an infection event or a recovery event. Every simulation continues until no infectious individuals can be found in the population.

\citet{keeling2005networks} provide a review of properties of epidemics over a variety of common network structures, including Erdős–Rényi random networks, lattices, small-world networks, scale-free networks, and spatial networks. Others~\citep{kiss2005disease,miller2009percolation} attempt to design disease spreading networks where node connections are clustered. On the other hand, other
works~\citep{Zhang2015modeling,liu2018measurability,Aleta2020} seek to construct networks from real-world contact data. In this work, we employ random networks with controllable epidemic parameters to analyze the limits of contact tracing and random testing strategies.

The number of secondary infections of individuals has been shown to be heterogeneous, and numerous ``superspreading events'' have been reported~\citep{shen2004superspreading}. \citet{Lloyd2005} proposed to model this phenomenon with a negative-binomial distribution parameterized by the basic reproduction number $R_0$ and a dispersion parameter $k$. The dispersion parameter $k$ controls the superspreading effect of a given disease, e.g., $k=0.16$ was estimated for SARS~\citep{Lloyd2005} where lower $k$ values indicate a stronger superspreading effect. Specifically, modeling efforts on the superspreading effect of the ongoing COVID-19 pandemic have been focused on estimating the dispersion parameter~\citep{endo2020estimating}, understanding the causes~\citep{Nielsen2020} and evaluating control strategies~\citep{kain2020chopping}.

As a tool for containing infectious epidemics, contact tracing is the process of identifying individuals who have been in contact with known infected cases. This process usually involves intensive manual efforts from specialists in interviewing infected patients and reconstructing their potential contacts~\citep{ferretti2020quantifying}. Prior work has targeted improving the efficiency and accuracy of this process through communication traces~\citep{farrahi2014epidemic} or dedicated digital apps~\citep{anglemyer2020digital,ahmed2020survey}.

Effective contact tracing is considered to be critical in controlling an ongoing epidemic~\citep{world2020contact,park2020contact}, which successfully contained SARS~\citep{lipsitch2003transmission}. However, this tool has experienced failures before, such as the British foot-and-mouth epidemic~\citep{ferguson2001transmission}. Therefore, a rich set of literature has studied the effectiveness of applying contact tracing in controlling epidemics~\citep{klinkenberg2006effectiveness,armbruster2007contact,huerta2002contact} and, especially, in the ongoing COVID-19 pandemic~\citep{ferretti2020quantifying,keeling2020efficacy,kretzschmar2020impact,Aleta2020,cebrian2021past}.

\section*{Results}\label{sec:results}

\subsection*{Superspreading random networks} The simplest random networks, the Erdős–Rényi (ER) networks~\citep{erdHos1960evolution}, have been shown to reduce the epidemic models to their classical fully mixed variants~\citep{keeling2005networks}. Therefore, in the context of epidemics, they can be fully characterized given the number of nodes $N$ and the $R_0$, $\beta$, and $\gamma$ parameters. In the same spirit, we have developed a new method to generate random networks taking into account the superspreading effect. Given the number of nodes $N$, the epidemic model parameters and the dispersion parameter $k$, we can fully define the node degree distributions of such networks, which we call superspreading random networks.

We show that the node degree distribution of a superspreading random network is defined as
\begin{equation}
    p(i \mid k, R_0, \beta, \gamma) = \frac{NB(i-1 \mid k, \frac{R_0 (\gamma + \beta)}{\beta})}{i\sum_{j=1}^{\infty}\frac{NB(j-1 \mid k, \frac{R_0 (\gamma + \beta)}{\beta})}{j}}
\end{equation}
where $NB(\cdot \mid k, \frac{R_0 (\gamma + \beta)}{\beta})$ is a negative binomial distribution parameterized by the dispersion parameter $k$ and mean $\frac{R_0 (\gamma + \beta)}{\beta}$. We then apply the configuration model~\citep{newman2003structure} to generate random networks.
The detailed derivation is explained in Methods.

\begin{figure*}[!tbp]
    \centering
    \begin{subfigure}{0.48\textwidth}
        \includegraphics[width=\textwidth,page=1]{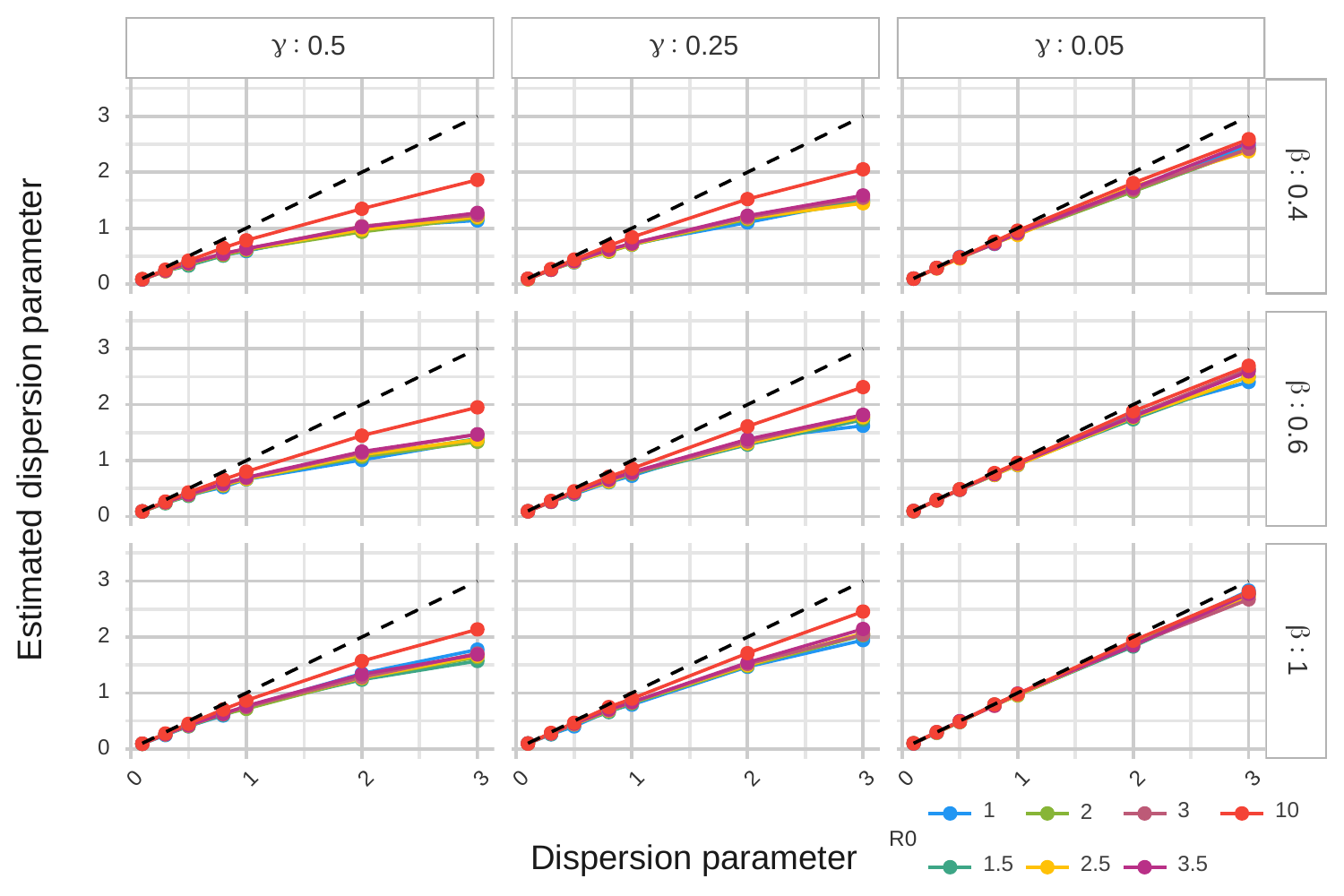}
        \caption{}
        \label{fig:dispersion_1}
    \end{subfigure}
    \hspace*{\fill}
    \begin{subfigure}{0.48\textwidth}
        \includegraphics[width=\textwidth,page=2]{images/compare_poisson_dispersion}
        \caption{}
        \label{fig:dispersion_2}
    \end{subfigure}
    \begin{subfigure}{0.99\textwidth}
    \includegraphics[width=\textwidth,page=1]{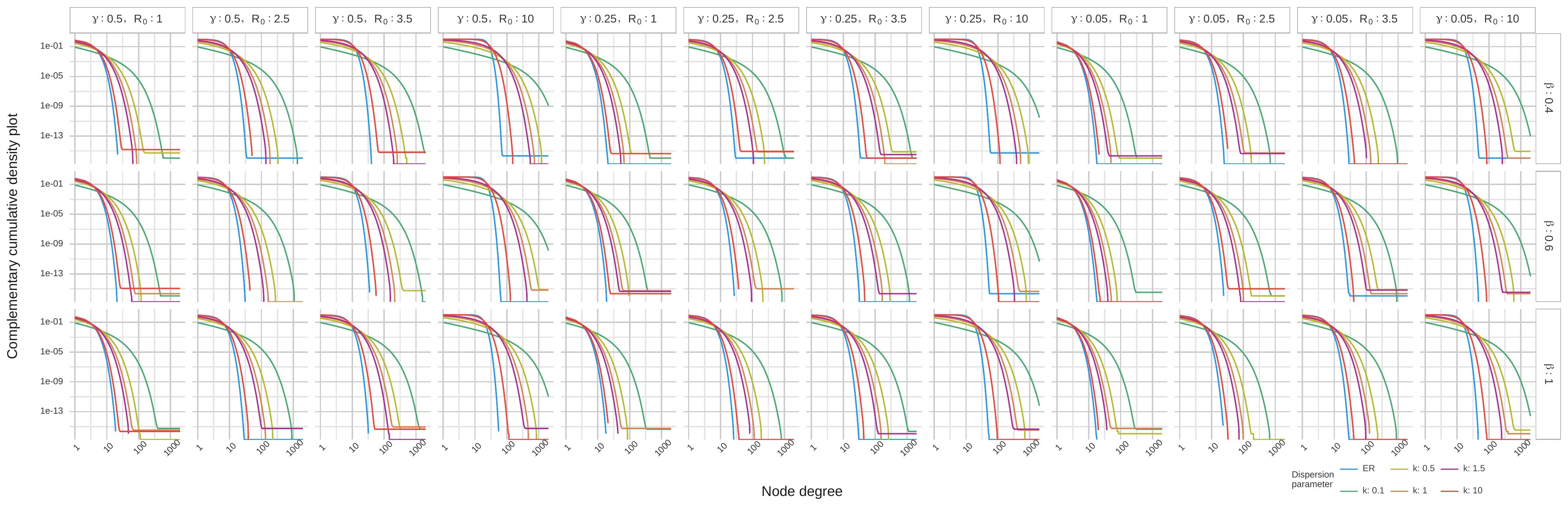}
    \caption{}
    \label{fig:dispersion_3}
    \end{subfigure}
    \caption{
        Comparison of network properties given different parameters. (a) shows the difference between chosen dispersion parameters and estimated dispersion parameters in simulations. (b) depicts the estimated dispersion parameters of ER random networks. (c) presents a complementary cumulative density plot of node degrees for dispersion networks with varying parameters.
    }
	\label{fig:network_properties}
\end{figure*}

Given the assumption of infection independence (see Methods), we seek to ascertain that desired dispersion parameters are achieved and maintained throughout the exponential phase of the epidemic. \cref{fig:dispersion_1} shows that this is the case up to $k=1$ regardless of the recovery and infection rates and that it degrades at lower infection rates and higher recovery rates. In contrast, in ER networks, dispersion parameters cannot be controlled. As shown in~\cref{fig:dispersion_2}, it is always above $1$, and, in general, it increases with higher infection rates and lower recovery rates. Both~\cref{fig:dispersion_1,fig:dispersion_2} show results for the first $100$ infected nodes with $10$ initial infected nodes. Similar results for higher numbers of infected nodes are shown in Supplementary~\cref{fig:network_properties}. As it can be seen, using ER networks for an epidemic in which $k$ is known to be low (such as COVID-19 or SARS) would lead to a misrepresentation of the distribution of the infections. 

Prior works~\citep{Lloyd2005,Nielsen2020} typically model heterogeneous secondary infection numbers with variances in individual infectiousness, i.e., posing a Gamma distribution on infection rates. Supplementary~\cref{fig:network_properties_2} shows that, for low $R_0$ values and especially for low recovery rates, the difference between the intended dispersion parameter and the achieved dispersion parameter in the first $100$ infected nodes is significant and much larger in this case than for the superspreading network. 

Finally, we show in~\cref{fig:dispersion_3} the node degree distribution of superspreading networks for different values of the dispersion parameters, $R_0$, infection rates, and recovery rates. Other network measures, such as clustering coefficients, are shown in Supplementary~\cref{fig:network_properties_clustering_coefficients}. 

\subsection*{A study of contact tracing and Random Testing}
See Methods for a complete description of the computational setup used throughout this section.
\subsubsection*{Correlation between positive rates and actual infections}
\begin{figure*}[!tbp]
    \centering
    \makebox[0.5\linewidth][c]{
    \begin{subfigure}{0.95\textwidth}
    \includegraphics[width=\textwidth,page=3]{images/SIR_positive_correlation_change_varying_parameters_selected_main_figure_beta_0.6}
    \caption{}
    \label{fig:complete_parameter_test_main_1}
    \end{subfigure}
    }
    \makebox[0.5\linewidth][c]{
    \begin{subfigure}{0.95\textwidth}
    \includegraphics[width=\textwidth,page=2]{images/SIR_positive_correlation_change_varying_parameters_selected_main_figure_beta_0.6}
    \caption{}
    \label{fig:complete_parameter_test_main_2}
    \end{subfigure}
    }
    \makebox[0.5\linewidth][c]{
    \begin{subfigure}{0.95\textwidth}
    \includegraphics[width=\textwidth,page=1]{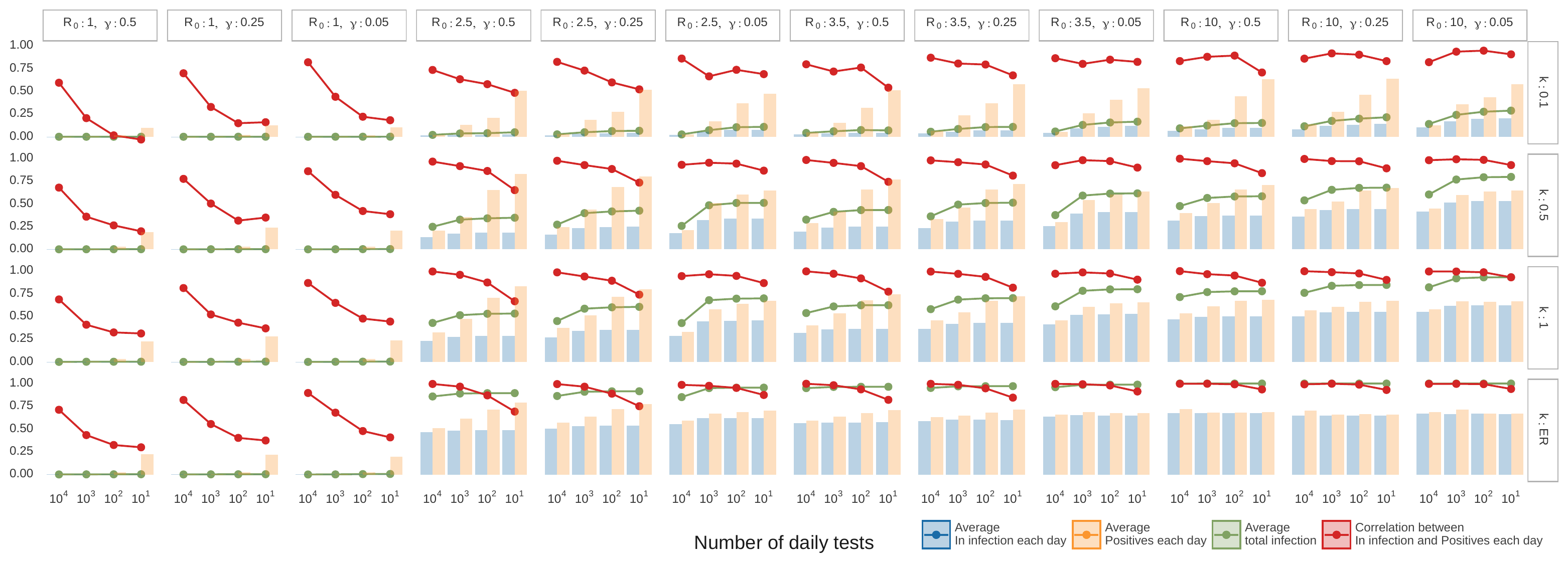}
    \caption{}
    \label{fig:complete_parameter_test_main_3}
    \end{subfigure}
    }
    \caption{
        Correlation plots between daily infections and positive rates for SIR models with different parameters, $\beta=0.6$ and $P_H=0.05$. (a) shows correlation plots for random testing, (b) shows correlation plots for forward contact tracing, and (c) shows correlation plots for backward contact tracing. Red lines represent correlation values and green lines indicate expected final total infections. Bars represent expected daily infection ratios and daily average positive rates. Note that bars are scaled by the maximum values.
    }
	\label{fig:complete_parameter_test_main}
\end{figure*}

Positive rates from testing are widely adopted for estimating underlying infection statuses in populations~\citep{wu2020substantial}. Therefore, in this part, we evaluate the correlation between positive rates (see Methods) and actual infected numbers on each day. \cref{fig:complete_parameter_test_main} depicts the averaged daily correlations for SIR model (A similar study is depicted in Supplementary~\cref{fig:complete_parameter_test_main_seir} for the SEIR model).

First, we can see that, in general, for small $R_0$, small $k$, large $\gamma$ and a small number of tests, the correlations are relatively poor, so positive rates from all strategies are misrepresenting the actual infection. These correlations become better as the $R_0$, $k$ and the number of tests increase.

For RT (\cref{fig:complete_parameter_test_main_1}), we observe close proportions between daily confirmed rates and actual infection rates in general. We can conclude that random testing, for a sufficiently large $R_0$ and number of tests, is a good estimator of the actual epidemic progress. Note that correlations always improve as the number of daily tests increases, even though the testing is also affecting the epidemic since positives are being quarantined. For a scenario in which testing does not affect the epidemic ($P_q=0$) see Supplementary~\cref{fig:complete_parameter_test_main_1_pq0} for SIR or Supplementary~\cref{fig:complete_parameter_test_main_1_pq0_seir} for SEIR.

\cref{fig:complete_parameter_test_main_2,fig:complete_parameter_test_main_3} show the correlations for forward and backward contact tracing, respectively. We can see that in both cases, positive rates often overestimate the actual infections, especially for low $k$ and $\gamma$ values. Moreover, BCT overestimates more than FCT, especially when the number of tests is low and the epidemic has a low dispersion parameter. This also means that BCT is better at finding positive contacts under these conditions. 

Also, we see instances in which CT daily positive rates increase when more daily tests are deployed, whereas actual infections in the population decrease. For example, when $R_0=10, \gamma=0.05, k=0.5$ in~\cref{fig:complete_parameter_test_main_2}, $1,000$ daily tests lead to higher proportions of positive cases than $100$ daily tests, which results in a false estimation of the epidemic. 

In conclusion, positive rates from contact tracing can be misleading and should be considered with caution. However, positive rates from random testing with a sufficiently large number of daily tests show more promise as a proxy for the actual epidemic progress (A similar trend can be seen in Supplementary~\cref{fig:complete_parameter_test_main_seir} for the SEIR model).

\subsubsection*{Analysis on final epidemic infections} 
\begin{figure*}[!tbp]
    \centering
    \begin{subfigure}{0.78\textwidth}
        \includegraphics[width=\textwidth,page=1]{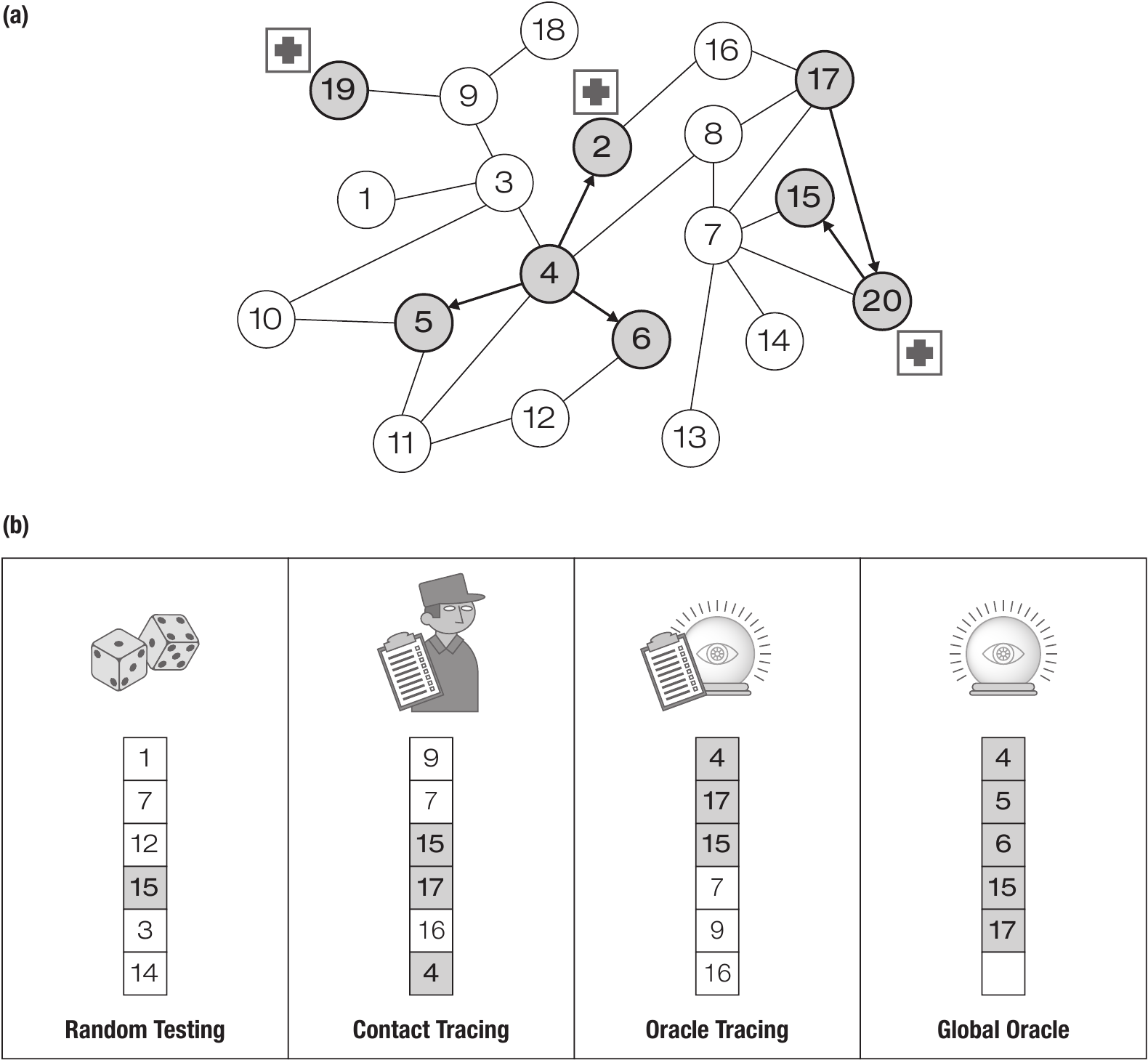}
    \end{subfigure}
    \caption{
        Cartoon representation of all contact tracing/testing strategies for a toy network. (a) shows a network where some nodes are shaded indicating that they are infected, and some of those include a cross symbol, indicating they are hospitalized, i.e., they are triggers of contact tracing. This network is a representation of a snapshot of the status of an epidemic in a given moment in time. (b) shows the four strategies considered for the toy network in (a): Random testing --- nodes are selected at random, the resulting list of nodes to test contains 1 infected individual that will test positive and be quarantined. Contact tracing --- contacts of all hospitalized nodes are included in the list in no particular order, we see that 3 infected individuals are in this list, although some non-infected nodes will be tested before them. Oracle tracing --- this is an ideal contact tracing strategy, where we assume there is an oracle to tell us beforehand who in the testing queue is infected and we prioritize testing these nodes. Global oracle --- in this ideal strategy we assume an oracle who tells us the actual infection status of the whole network so we only add infected nodes to the testing queue. Note that the differences between forward and backward contact tracing cannot be shown in a single day snapshot of the epidemic. Note that the number of positive nodes found for each strategy is dependent on the number of tests available per day. In this example, if the test limit were 6, the positives found would be 1,3,3 and 5, respectively; if the limit were 5, the positives found would be 1,2,3 and 5, respectively.
    }
	\label{fig:teaser_figure}
\end{figure*}

Here, in order to classify different epidemics, we introduce two novel strategies that represent both the ceiling of contact tracing and the ceiling of testing. \cref{fig:teaser_figure} shows a cartoon representation of the effect of these strategies on a toy infection network and their differences with both random testing and contact tracing. We call these ideal strategies oracles, and they are defined in Methods.

\begin{figure*}[!tbp]
    \centering
    \makebox[1\linewidth][c]{
    \begin{subfigure}{0.8\textwidth}
    \includegraphics[width=\textwidth,page=1]{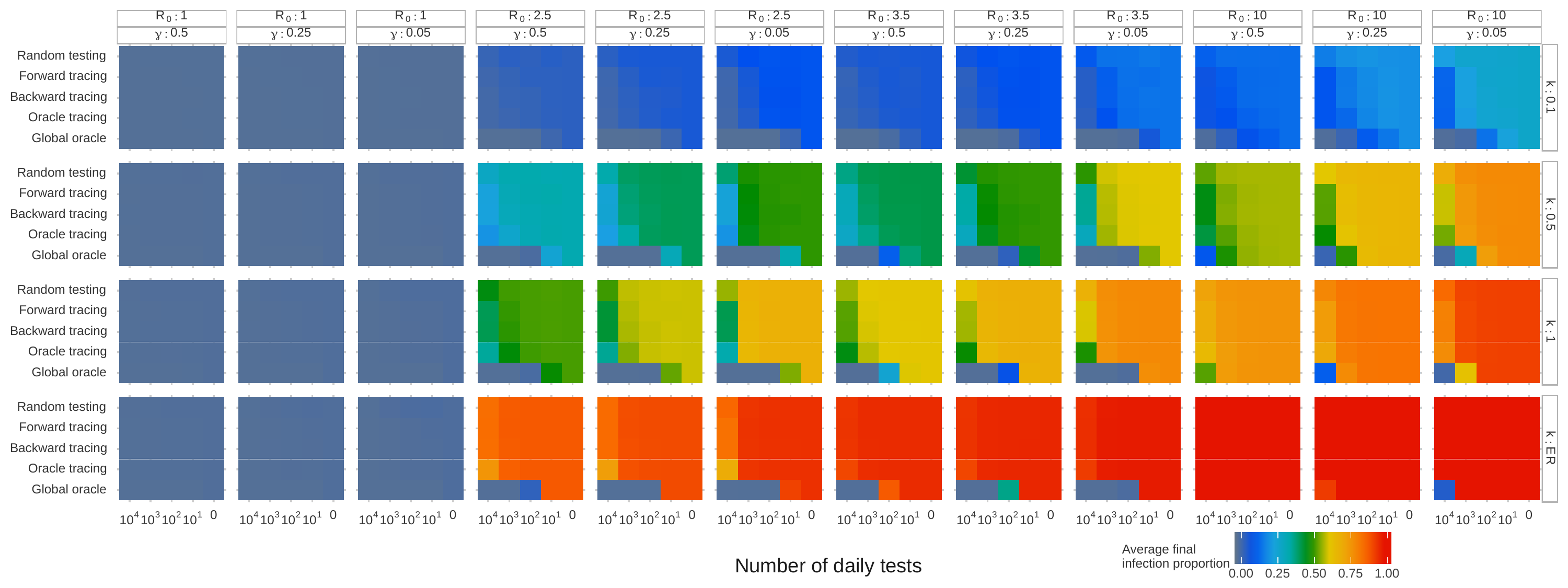}
    \caption{}
    \label{fig:complete_parameter_test_main_total_infection_1}
    \end{subfigure}
    }
    \makebox[1\linewidth][c]{
    \begin{subfigure}{0.8\textwidth}
    \includegraphics[width=\textwidth,page=3]{images/SIR_final_infection_and_days2end_main_figure_beta_0.6}
    \caption{}
    \label{fig:complete_parameter_test_main_total_infection_2}
    \end{subfigure}
    }
    \makebox[1\linewidth][c]{
    \begin{subfigure}{0.8\textwidth}
    \includegraphics[width=\textwidth,page=2]{images/SIR_final_infection_and_days2end_main_figure_beta_0.6}
    \caption{}
    \label{fig:complete_parameter_test_main_total_infection_3}
    \end{subfigure}
    }
    \caption{
        Evaluation of epidemic simulations over a set of parameters for SIR model with $\beta=0.6$ and $P_H=0.05$. (a) shows average final infections, (b) shows average final infections in top 5 communities and (c) shows average days to the last infections.
    }
	\label{fig:complete_parameter_test_main_total_infection}
\end{figure*}

When comparing average total infections, \cref{fig:complete_parameter_test_main_total_infection_1} shows that, for the networks simulated with $k=0.1$, around $20\%$ of the population may be infected in the worst scenario without any intervention and less than $10\%$ of population when $R_0 < 2.5$. 
On the other hand, when intervention operations are applied, epidemics with higher $\beta$ and $R_0$ are more difficult to contain. Only GOT contains epidemics within a small number of daily tests, which becomes more difficult as $R_0$ grows. COT performs slightly better than the other contact tracing strategies (forward and backward). However, BCT does not show obvious advantages over FCT. In general, contact tracing does not seem to have a considerable impact in the course of the epidemic, and much less so random testing. Also, note that all epidemics result in much larger infected populations when simulations occur over an ER network. For SEIR models in Supplementary~\cref{fig:complete_parameter_test_main_total_infection_seir}, it can also be found that incubation periods (when individuals are not contagious but they are detectable) result in more containable epidemics when compared to the same parameter sets in SIR models.

The nature of the superspreading events when $k$ is low renders a lower portion of the population infected at the end of the epidemic (Figure 3a). However, since superspreading events will have more impact in denser communities (connected components in the network), especially in those where initially infected individuals were found, we explore the average percentage of the infected population in the top $5$ largest communities in the network. Figure 3b shows that this has no effect for large $k$ values or in ER networks but shows noticeable differences, especially for $k=0.1$. Note that, for example, $R_0=2.5$ and $\gamma=0.05$ with no interventions reaches only $10$ to $15$ percent of infected individuals at the end of the epidemic, but around $80$ percent within the densest communities.

From \cref{fig:complete_parameter_test_main_total_infection_3} and Supplementary~\cref{fig:complete_parameter_test_main_total_infection_3_seir}, we see that high $k$ and small $\gamma$ lead to longer epidemic diffusion times, peaking at $R_0=2.5$. Here we see an opposite trend to~\cref{fig:complete_parameter_test_main_total_infection_1} with longer times for smaller $R_0$ and lower $k$ values. Also, contact tracing seems to have a greater impact on times than on the total infected population. When comparing SEIR models to SIR models, longer days are observed overall due to the incubation periods.

Finally, \cref{fig:complete_parameter_test_main_total_infection} helps us classify parameter sets in terms of how containable they are. Thus, we can see that epidemics with $R_0 = 1$ can be contained even without any interventions; however, contact tracing does have an impact in the densest communities and in the final time of the epidemic. For $k=0.1$ tracing interventions are capable of partially containing the epidemics with a sufficient number of resources. For larger $k$ values and as $R_0$ increases, epidemics are not containable with RT or any CT strategies. Only the global oracle, equipped with large numbers of daily tests, can have an impact. Lastly, for extreme $R_0$ values (10 and beyond) and large $k$ values (above $1$ and for ER networks), there is no intervention that can contain the epidemic.

\subsubsection*{A mixed strategy for containment and surveillance}

Many countries have established daily risk level thresholds that are accompanied by different measures. For illustration purposes, we choose the levels defined by the Spanish government (see Methods).

\begin{figure*}[!tbp]
    \centering
        \begin{subfigure}{0.65\textwidth}
        \includegraphics[width=\textwidth,page=4]{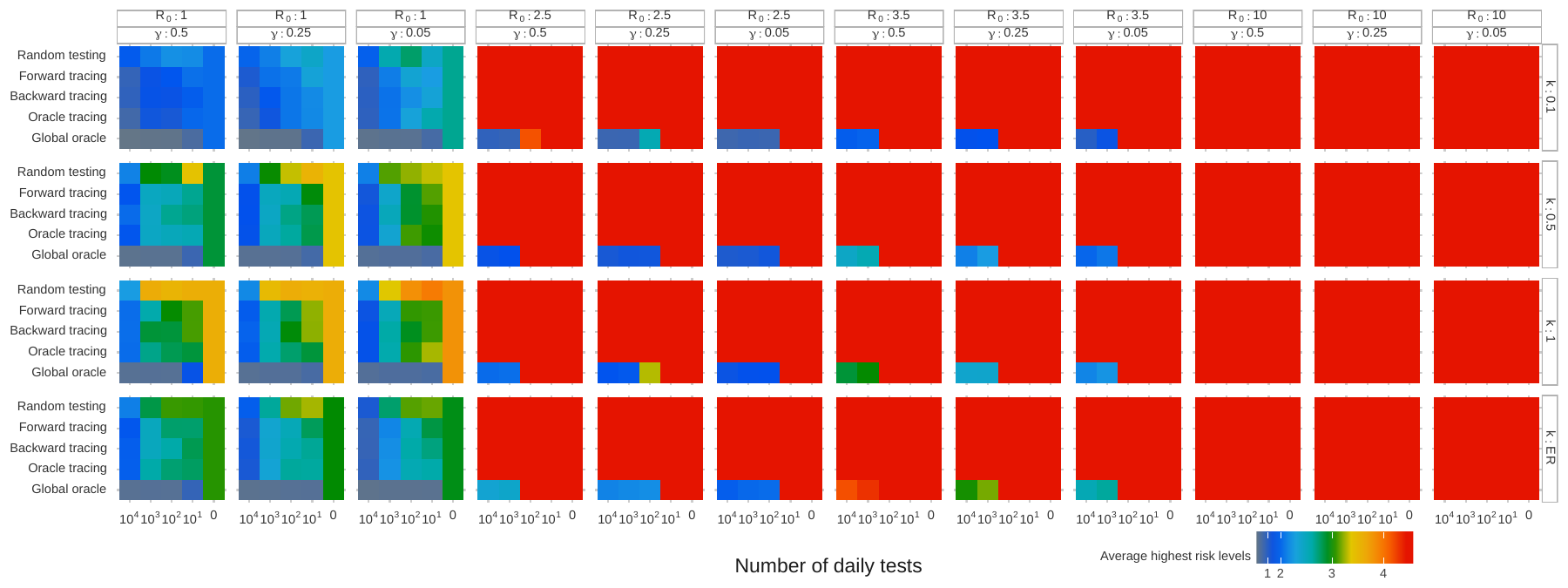}
        \caption{}
        \label{fig:complete_parameter_test_main_risk_levels_1}
        \end{subfigure}
        \begin{subfigure}{0.65\textwidth}
            \includegraphics[width=\textwidth,page=5]{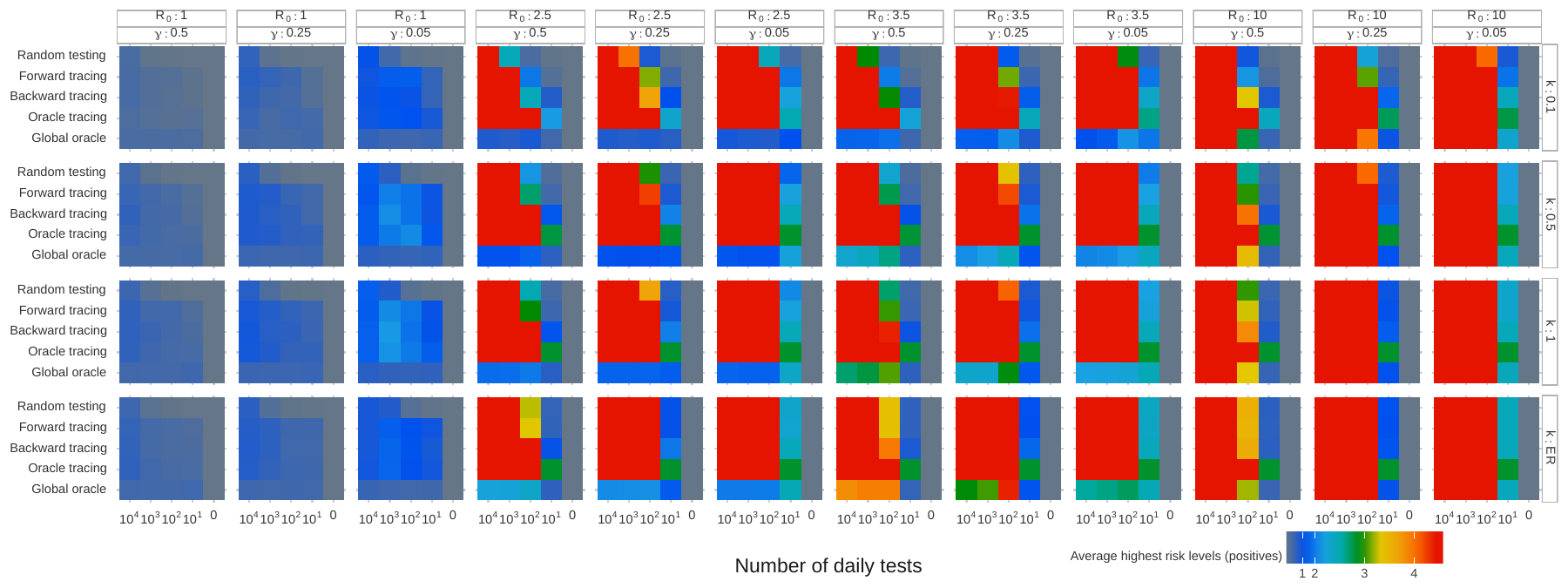}
            \caption{}
            \label{fig:complete_parameter_test_main_risk_levels_2}
        \end{subfigure}
        \begin{subfigure}{0.65\textwidth}
            \includegraphics[width=\textwidth,page=6]{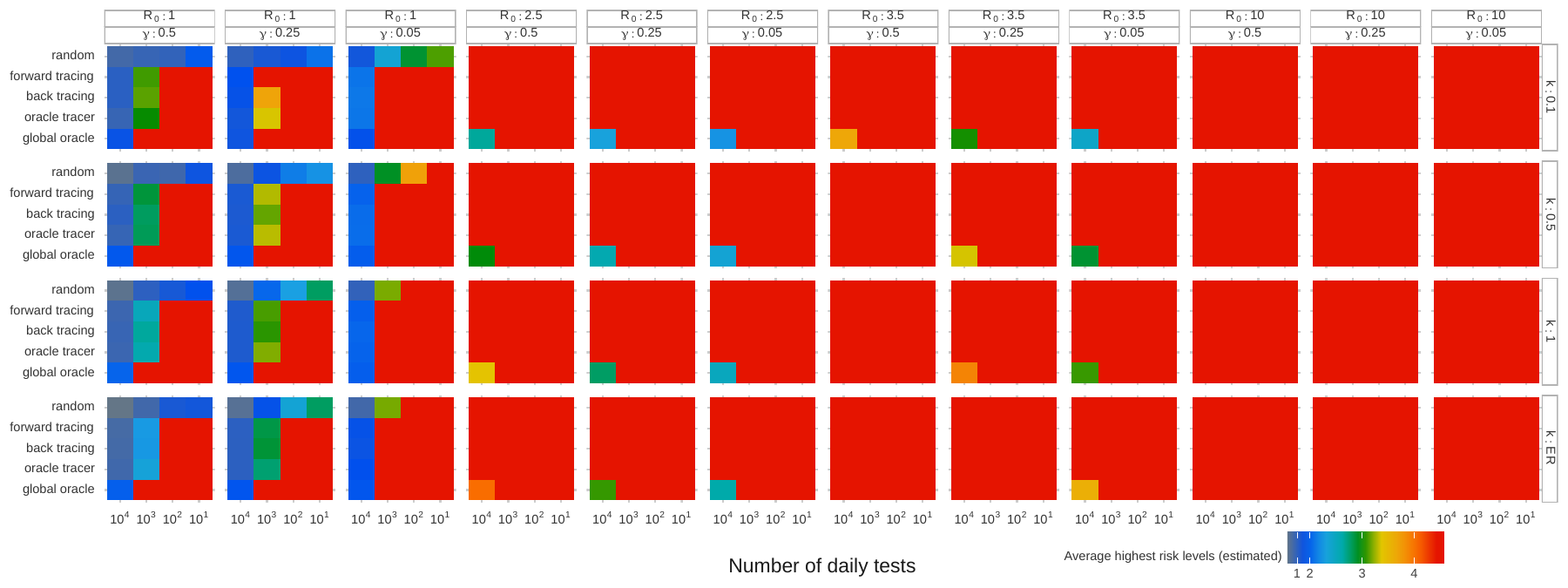}
            \caption{}
            \label{fig:complete_parameter_test_main_risk_levels_3}
        \end{subfigure}
        \begin{subfigure}{0.65\textwidth}
            \includegraphics[width=\textwidth,page=7]{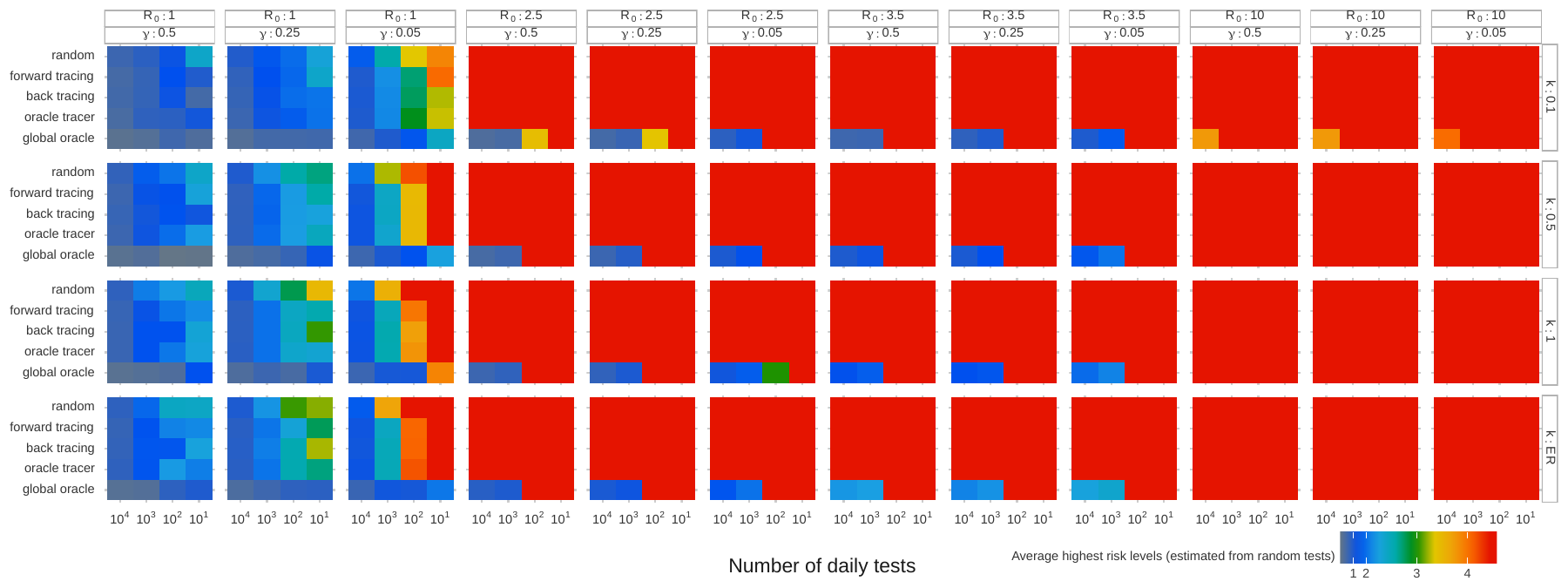}
            \caption{}
            \label{fig:complete_parameter_test_main_risk_levels_4_RT_100}
        \end{subfigure}

    \caption{
        Summary of threat levels of SIR model simulations given different epidemic parameters and intervention strategies where $\beta=0.6$ and $P_H=0.05$. (a) shows actual highest risk levels for SIR models, (b) shows highest threat levels from confirmed cases, (c) shows highest threat levels from positive rates and (d) shows highest threat levels for the mixed strategy (positive rates estimated only from random tests).
    }
	\label{fig:complete_parameter_test_main_risk_levels}
\end{figure*}

\cref{fig:complete_parameter_test_main_risk_levels} depicts the maximum threat level achieved during an epidemic for each of the epidemic parameter values, the number of daily tests, and different contact tracing/testing strategies.

\cref{fig:complete_parameter_test_main_risk_levels_1} shows the actual levels achieved. As it can be seen, for almost all cases, the maximum threat level is reached at some point in time during the epidemic (except for low $R_0$ values). \cref{fig:complete_parameter_test_main_risk_levels_2} shows the maximum threat level when considering all positive tests as the actual total positives in the population. This inference consistently underestimates the maximum threat level. On the other hand, \cref{fig:complete_parameter_test_main_risk_levels_3} shows the maximum threat levels if those are inferred from the positive rates found during testing. As expected from our correlations section results, this scenario consistently overestimates threat levels (except RT).

Finally, 
we constructed a mixed strategy in which $100$ daily tests were devoted to random testing regardless of the strategy (for $10$ daily tests we devoted $5$ to random testing and for $100$ daily tests $50$ were devoted to random testing). \cref{fig:complete_parameter_test_main_risk_levels_4_RT_100} shows the maximum threat levels reached with this new strategy when only positive rates from random testing are used. Although far from perfect, we can see that this strategy shows maximum threat levels very similar to the actual ones shown in~\cref{fig:complete_parameter_test_main_risk_levels_1}. 

Note that this mixed strategy could potentially result in different infection levels, since not all tests are devoted to the given strategy. However, Supplementary~\cref{fig:complete_parameter_test_main_total_infection_rt100_seir} shows that the differences between this strategy and the non-mixed ones (compare with~\cref{fig:complete_parameter_test_main_total_infection}) are negligible.

\subsubsection*{Known diseases}
As an example and proof of concept, we seek to quantify the effect of intervention strategies on several known diseases with their reported epidemic parameters, including Measles~\citep{stone2000theoretical,antona2013measles,nishiura2017assessing}, H1N1~\citep{furushima2017estimation,koliou2009epidemiological,dorigatti2012new}, Ebola~\citep{lekone2006statistical,althaus2015ebola}, SARS~\citep{chowell2003sars} and COVID-19~\citep{Aleta2020,endo2020estimating}. Their detailed parameters are listed in Supplementary~\cref{tab:exhaustive}. In our simulations, we fix the population number to $N=100,000$ and $I_0=10$ initial infections. Again, we vary the number of daily tests from $0$ to $10,000$.

\begin{figure*}[!tbp]
    \centering
    \begin{subfigure}{0.496\textwidth}
    \includegraphics[width=\textwidth,page=1]{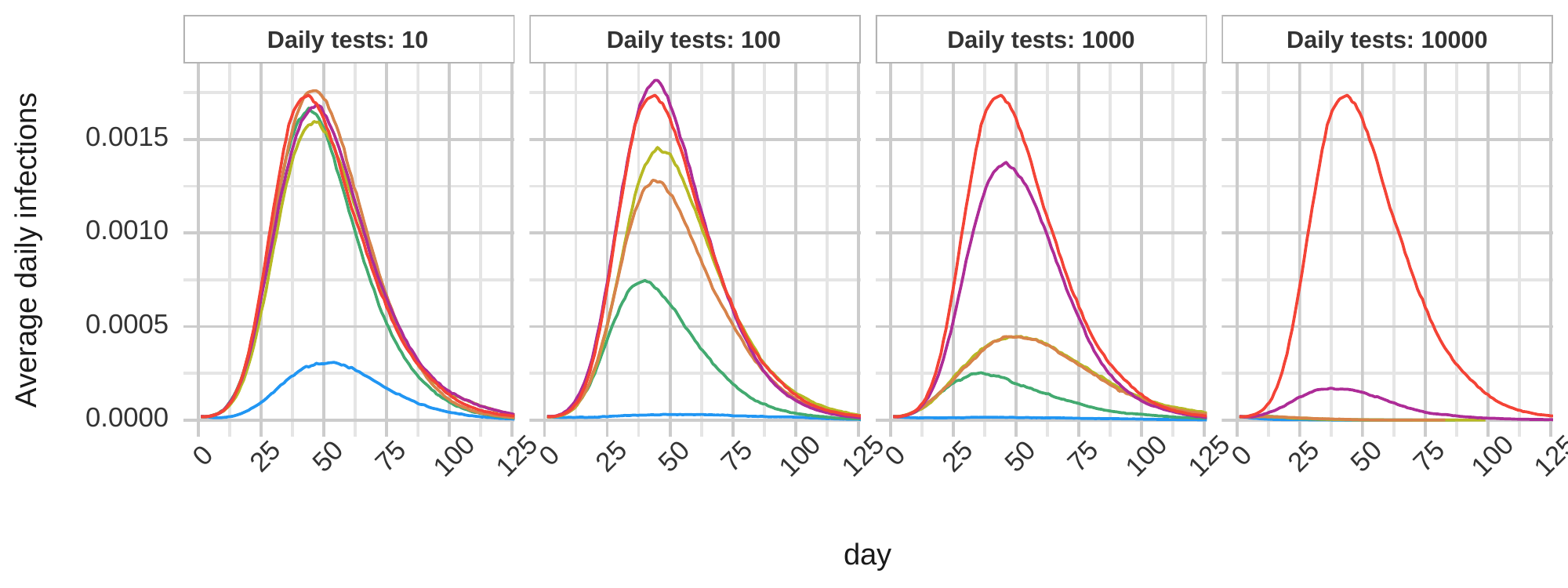}
    \caption{COVID-19}
    \label{fig:known_diseases_1}
    \end{subfigure}
    \begin{subfigure}{0.496\textwidth}
    \includegraphics[width=\textwidth,page=2]{images/known_diseases_infections_tmp_RT_100_2_nbinomial}
    \caption{Ebola}
    \label{fig:known_diseases_2}
    \end{subfigure}
        \begin{subfigure}{0.496\textwidth}
    \includegraphics[width=\textwidth,page=3]{images/known_diseases_infections_tmp_RT_100_2_nbinomial}
    \caption{SARS}
    \label{fig:known_diseases_3}
    \end{subfigure}
        \begin{subfigure}{0.496\textwidth}
    \includegraphics[width=\textwidth,page=4]{images/known_diseases_infections_tmp_RT_100_2_nbinomial}
    \caption{H1N1}
    \label{fig:known_diseases_4}
    \end{subfigure}
        \begin{subfigure}{0.496\textwidth}
    \includegraphics[width=\textwidth,page=5]{images/known_diseases_infections_tmp_RT_100_2_nbinomial}
    \caption{Measles}
    \label{fig:known_diseases_5}
    \end{subfigure}
    \caption{
        The average daily infection ratios of five known diseases by varying number of daily tests, under different intervention strategies.
    }
	\label{fig:known_diseases}
\end{figure*}

\cref{fig:known_diseases} depicts the epidemic curve of the simulation results of the five epidemics under different interventions. Supplementary~\cref{fig:known_diseases_cum} shows the same results for the cumulative cases.

In \cref{fig:known_diseases_1,fig:known_diseases_2,fig:known_diseases_3,fig:known_diseases_4}, overall, the operations with oracle strategies contain epidemics better, while random testing requires much more daily tests to control the epidemic spreading. In particular, both contact tracing operations (forward and backward) lead to similar average daily infection curves.
When comparing different diseases, SARS (\cref{fig:known_diseases_3}) is the most containable epidemic, while Measles cannot be controlled due to its extreme $R_0$ value and its infection processes in all simulations finished shortly compared to other epidemics as shown in~\cref{fig:known_diseases_5}. These two are the extreme points where an epidemic can be controlled with CT and where it cannot be controlled even with GOT, respectively. For the rest, H1N1 (\cref{fig:known_diseases_3}) can be contained with relatively few resources, and Ebola (\cref{fig:known_diseases_2}) and Covid-19 might need one order of magnitude more to reach containment. Note that SEIR models are easier to contain under the assumption that E individuals are not contagious, but they will be tested positive. Finally, note that the global oracle can control all epidemics with few resources, except for Measles.

Supplementary~\cref{fig:known_diseases_poisson} and Supplementary~\cref{fig:known_diseases_cum_poisson} show the same results when we disregard the dispersion parameter and run the simulations over ER networks, both for the epidemic curve and cumulative cases, respectively. It can be seen that final infections are much higher than those found in superspreading networks with low dispersion parameters.

Of relevance, with the parameters for COVID-19 from~\citep{Aleta2020,endo2020estimating}, only around $10\%$ of the population is infected. A reasonable amount of daily tests can reduce this proportion to around $2-5\%$. This is mainly due to the low dispersion parameter and the static nature of our network simulations (and no re-introductions). We believe this is in accordance with what most countries are suffering nowadays in terms of infected population per wave~\citep{knockreport,sevillano_2020,last2020first}.

Finally, Supplementary~\cref{fig:known_diseases_risk} shows the daily threat levels for each of the epidemics. Note that the maximum threat level is reached at some point during the epidemic in all cases (even when the final proportion of infected individuals is low -- see Supplementary~\cref{fig:known_diseases_cum}). As expected, inferring threat levels from positives found in tests underestimates the actual infections, whereas inferring threat levels from positive rates from all tests overestimate them. In general, inferring threat levels from positives found only in random testing (mixed strategy) is a much closer estimate to the actual levels. 
\section*{Discussion}
In this paper, we study the computational bounds of contact tracing and random testing. 
We introduce random networks that take into account the superspreading effect with controllable dispersion parameters for the first time. We first find that backward contact tracing is slightly better than forward contact tracing for low dispersion parameters and a small limit of daily tests. We then find the limitation of contact tracing as a means to describing the actual epidemic status. Afterward, we provide a classification of epidemics in terms of how containable they are, in which we find that there is a gap between classic contact tracing and optimal contact tracing, and between this an optimal testing. The implications are exploring both smart contact tracing and smart testing techniques is worth it. We also see that the length of the epidemic can be misleading and that contact tracing also has an impact on it.

Finally, we want to address the fact that other recent works on contact tracing models~\citep{Aleta2020,kretzschmar2020impact,ferretti2020quantifying} for COVID-19 show more optimistic results than what transpires from our study regarding the impact of contact tracing. We believe three main factors contribute to this difference: (1) The superspreading effects controlled by the dispersion parameter are not explicitly considered in these works, while our study is done over networks where we control the dispersion parameter for the simulations. (2) Testing and/or tracing resources are unlimited for these references, while we have analyzed various levels of available testing or tracing resources up to $10\%$ of the total population of daily tests. (3) Lastly, most of these papers (~\citep{Aleta2020,ferretti2020quantifying}) consider between $35\%$ and $70\%$ of infected individuals as triggers of contact tracing, while for this work, only hospitalized individuals (which is $5\%$ of infected individuals throughout the main text) and positives found via random testing or contact tracing are considered.

For illustration purposes, we turn our focus to the exhaustive work carried out in~\citep{Aleta2020} to try to ascertain objectively whether these three issues can explain the differences observed. First, we see that, when we run the epidemic diffusion process as stated in~\citep{Aleta2020} (see Methods) over the network provided in \url{https://github.com/aaleta/NHB\_COVID}, the resulting dispersion parameter is $k=2.5$, which is higher than the estimated value of $0.1$ (\citepAP{endo2020estimating}). Second, again with the same parameters from~\citep{Aleta2020}, Supplementary~\cref{fig:aleta_model_1} shows the number of tests performed each day. It can be seen that the testing peak goes over $500$ tests (for $I_0=1$), which represents over $5\%$ of the total population of daily tests to achieve a final $91\%$ of the infected population. Finally, Supplementary~\cref{fig:aleta_model_2} brings~\citep{Aleta2020} framework closer to our own by limiting the number of daily tests and setting contact tracing trigger cases to hospitalizations (with a probability of $2\%$). Here we can see how the impact of contact tracing is drastically reduced. 

In conclusion, our study provides a cautionary tale of contact tracing. It also suggests a separation between tracing as a means to containing an epidemic and testing as a means of inferring the progress of the epidemic. Moreover, it highlights the need for more intelligent testing strategies in order to contain most of the possible (future) epidemics.

\clearpage

\section*{Methods}\label{ap:models}

\subsection*{Contact networks simulation}

\subsubsection*{Simulate Erdős–Rényi random networks}
\citet{bartlett2012epidemic} reveal the analytical connection between the epidemic parameters and network parameters for Erdős–Rényi random networks~\citep{erdds1959random}, i.e., 
\begin{align}\label{eq:r_0_network}
    R_0=\frac{\beta(N-1)K}{\gamma + \beta}
\end{align}
where a network is parameterized by the average node degree $K$. When $N\rightarrow \infty$, the node degrees of these networks are Poisson distributed with a mean value $\frac{R_0 (\gamma + \beta)}{\beta}$.  

\subsubsection*{Simulate networks with dispersion parameters}\label{sssec:dispersion_network}
In this section, we describe the algorithm for simulating random contact networks that lead to negative binomial distributed secondary infections with given parameters.

To generate a random network, we first seek to derive its node degree distribution. We provide here some definitions including several probability generating functions (PGFs) following~\citep{newman2002spread}:
\begin{itemize}
    \item Given a random network with its degree distribution $\Prob(d=i) = p_i$, its PGF is $G_0(x)=\sum_{i=1}^{\infty} p_i x^i$ and the average degree equals $ <i> =G_0'(x=1)= \sum_{i=1}^{\infty} i p_i $.
    \item The \textit{excess edge} of a vertex is defined as the number of remaining edges connected to the vertex when follow a random edge to that vertex. This corresponds the remaining neighbor counts when disease diffuses to a new individual. The probability that a vertex at the end of a random edge has excess degree $i-1$ is given in~\citep{newman2002spread} as $\frac{ip_i}{<i>}$. Therefore, the PGF for the excess degree of a vertex is
    \begin{equation}
        G_1(x) = \frac{\sum_{i=1}^{\infty}ip_i x^{i-1}}{\sum_{i=1}^{\infty}i p_i}
    \end{equation}
    \item More importantly, given the probability that a infected individual infects his/her neighbor $T$, the PGF of the number of infected neighbors is
    \begin{equation}
        G_0'(x) = G_0(1+(x-1)T)
    \end{equation}
    Similarly, the PGF for the excess occupied degree is $G_1'(x)=G_1(1+(x-1)T)$
\end{itemize}

We are then able to derive the degree distribution. The assumption of a negative binomial (NB) distributed secondary infections indicates that $G_1'(x)$ is also the PGF of an NB distribution. Given the average secondary infection $R_0$ and the dispersion parameter $k$, we have
\begin{align}
    G_1'(x) &= \sum_{n=0}^{\infty} \binom{n+k-1}{n} \left(\frac{R_0}{R_0 + k}\right)^n \left(\frac{k}{R_0 + k}\right)^k x^{n} \\
    &= \left(1 + \frac{R_0}{k}(1-x) \right)^{-k}
\end{align}
where the second step is due to the binomial theorem. With a change of variable, we get
\begin{align}
    G_1(y) = \left(1 + \frac{R_0}{Tk}(1-y) \right)^{-k}
\end{align}
where $G_1(y)$ is simply another NB with mean $\frac{R_0}{T}$ and dispersion parameter $k$. We denote its probability mass function as $q_{i-1}$ which, as mentioned above, is equal to $\frac{ip_i}{<i>} = q_{i-1}$. Due to $\sum_{i=1}^{\infty} p_i = 1$, we are able to compute the average node degree
\begin{align}
    <i> &= \frac{1}{\sum_{i=1}^{\infty} \frac{q_{i-1}}{i}} \\
     &= \frac{\frac{R_0}{T} (1-k)}{(\frac{kT}{kT + R_0})^k (k+\frac{R_0}{T})-k}
\end{align}
which can be easily solved numerically.

We then need to define the infection probability $T$ given the infection rate and recovery rate. For an infected individual, assuming his/her recovery follows a rate $\gamma$ and he/she is infecting a neighbor with a rate $\beta$, we can easily derive the probability of the neighbor being infected as $\frac{\beta}{\gamma + \beta}$. If we further introduce a relaxed assumption that all neighbors are infected i.i.d., we then have $T=\frac{\beta}{\gamma + \beta}$.

Given the derived degree distributions, we can then simulate a random network by applying a configuration model~\citep{newman2003structure}. We note that any self-loops or parallel edges are removed from the generated networks.

\subsubsection*{Explore empirical dispersion parameters in simulations}
Contact networks simulated via the method described in~\nameref{ap:models} is based on an assumption that one infectious individual infects his/her neighbors independently. However, this relaxed assumption may lead to errors between chosen dispersion parameters and their empirical values in simulations. Here we explore such discrepancies via simulations.

\subsubsection*{Clustering coefficients}
\citet{sriram12} show that clustering coefficients of the configuration model are defined as:
\begin{equation}
    \frac{1}{N}\frac{(<i^2> - <i>)^2}{<i>^3}
\end{equation}
where $N$ is the number of nodes. This indicates that the clustering coefficients are $0$ in the limit of large networks. 

\subsection*{Experimental setup}\label{sec:exhaustive}
Here we detail the main concepts of our experimental setup. As it will be seen, all modeling decisions are geared towards a more favorable scenario for the impact of contact tracing rather than towards a more realistic one. Therefore, the following results are best-case scenarios for all strategies. The main modeling considerations are as follows:

\begin{itemize}
    \item \textbf{Compartmental models.} We simulate outbreaks with two epidemic models, SIR and SEIR. The SIR model assigns three possible infection statuses to individual nodes, susceptible ($S$), infected ($I$), and recovered ($R$), whereas the SEIR model further introduces the exposed ($E$) for modeling incubation periods of diseases. While most nodes are in the susceptible status ($S$), $I_0$ number of nodes are initialized as infectious nodes ($I$) at $t=0$, who spread the disease to their neighbor nodes at a rate $\beta$. The dynamics of infectious periods (i.e., from $I$ to $R$) and incubation periods (i.e., from $E$ to $I$) of individual nodes are defined by rates $\gamma$ and $\kappa$, respectively. Besides, we introduce a hospitalized ($H$) compartment in~\nameref{sec:exhaustive} section to model the possibility of infected individuals disclosing their statuses via hospitalization. The rate of $I$ to $H$ is $\eta$, and it can be directly calculated from the probability of hospitalization $p_H$. Moreover, the basic reproduction number, $R_0$, is an important epidemic quantity that defines the average secondary infections caused by a single infected individual. Throughout the paper, results for SIR models are shown in the main text figures, while those for SEIR are depicted in supplementary figures. The main decision about the SEIR model is that individuals who are in incubation periods cannot infect but can be detected as positive in a test. Both models are extended with a Hospitalization rate, where some infected individuals become hospitalized. Hospitalized individuals are considered as automatic positives who trigger contact tracing.

    \item \textbf{Tracing and testing setups.} In our experiments, we explore the limits of contact tracing and random testing in containing epidemics. Given a fixed number of tests per day, these strategies are applied daily in epidemic simulations, and different individuals are then tested for their infection statuses. We assume the test results are provided immediately without false positives and false negatives. We disregard the number of tracers and assume instead the limiting resources are the daily tests. Calculating the number of tracers to daily tests is not discussed in this work.

    \item \textbf{Random testing (RT).} This operation devotes resources to randomized testing within a population. Random tests can be performed in real-life scenarios by contacting random people and asking them to take a test or by announcing voluntary tests at specific locations. Here, we assume all infection statuses will receive an equal probability of selection for tests and also that recovered individuals can be chosen for such random tests (only individuals that have tested positive and still have not recovered are not selected). These two decisions together allow us for a unique implementation that accounts for both real-life scenarios at the same time.

    \item \textbf{Contact tracing (CT).} During simulations, we maintain a queue of individuals to be tested and traced. Neighbors of positive cases, found by contact tracing, random testing, or hospitalization, are recorded in the queue. The probability of discovering a contact of an infected node is denoted as $P_c$ and is set to $1$ (akin to using a perfect contact tracing app~\citep{ahmed2020survey}). We note that, when the queue becomes empty, all remaining testing resources are devoted to random testing, so that all tests are performed every day. Two possible strategies arise when individuals in the queue are prioritized differently: 
    \textbf{Forward contact tracing (FCT)} orders the tests for individuals according to the times when they are added to the queue. On the other hand, \textbf{backward contact tracing (BCT)} prioritizes the tests to neighbors of positive cases, up to one hop in networks. This latter approach has been proposed for epidemics with low dispersion parameter~\citep{hethcote1984gonorrhea,muller2000contact,endo2020implication}, aiming to find sources of the infections.

    \item \textbf{Quarantine.} We assume a scenario where nodes that have tested positive are quarantined---i.e., removed from the contact network---with a probability that we denote as $P_q$. This probability is set to $1$.

    \item \textbf{Parameter exploration.} We quantify the effect of different intervention operations on a wide range of possible epidemics by varying all epidemic parameters. We also explore the space of initial infected (with no re-introductions) and the number of daily tests. A complete table of the parameters tested is presented in Supplementary~\cref{tab:exhaustive}. In the main text, we discuss a smaller set of parameters that are representative of the observed trends. Complete results for all parameters can be found in Supplementary Data 1.

    \item \textbf{Simulation setups} Given an epidemic model with chosen parameters, we first simulate $15$ different superspreading random networks based on the different network parameters but with a fixed number of nodes $N=100,000$ (some results with $N=1,000,000$ are shown in Supplementary Data 2). On each superspreading network, we randomly select $I_0$ individuals as initial infections and simulate an epidemic with applications of all intervention strategies daily. This process is repeated $30$ times on every contact network, which results in $450$ simulations in total. In all results that follow, the number of initially infected nodes $I_0$ is set to 10. Results for larger numbers of initially infected nodes are shown in Supplementary Data 1.
\end{itemize}

\subsection*{Positive rates from tracing/testing}

Since we cannot select (for testing) individuals that have already tested positive but are not yet recovered (which we call $ctd$ below), the positive rates are calculated as follows:
\[
Pos_{rate} =  ((positives/tests)*(N-ctd)+ctd)/N
\]

\subsection*{Oracles}
The \textbf{contact tracing oracle (CTO)} prioritizes individuals in the testing queue so that the actual infected ones are visited (tested) first. This provides an upper bound on the best contact tracing strategy.

The \textbf{global oracle testing (GOT)} assumes the availability of information about all newly infected individuals who are then tested and quarantined. Thus, the contact tracing queue is filled only with infected individuals. This assumption leads to an upper bound for future smart testing strategies.

We note that, although these two oracle scenarios are unrealistic in practice, they provide ideal upper-bounds for potential new strategies that might arise in the future.

\subsection*{Threat Levels}

We use the Spanish Government threat levels, which are based on the total infection number, in the last $14$ days, per $100,000$ individuals. These levels are the following: (1) less than $25$ infected, (2) between $25$ and $50$, (3) between $50$ and $150$, (4) between $150$ and $250$ and (5) over $250$ infected individuals per $100,000$ individuals. In practice, these levels also take into account positive rates and ICU occupancy, however, we ignore these here for the sake of simplicity.

\subsection*{Baseline Simulations from the Prior Work}
We apply the network provided in \url{https://github.com/aaleta/NHB\_COVID} by~\citet{Aleta2020}. The provided network is unweighted and has $10,000$ nodes and an average node degree of $10$. In terms of the simulations, we implement the epidemic parameter sets following the statistics detailed in~\citep{Aleta2020} where a $50\%$ probability of the discovery of symptomatic infections is applied and $40\%$ of contacts are successfully traced. The only two differences with ~\citep{Aleta2020} are: testing and quarantining are performed immediately (whenever tests are available) and no household contacts are automatically quarantined (since these are unknown in the network provided).

\subsection*{Software Implementation}
The simulation algorithm was implemented in Python based on the sample code provided by~\citet{kiss2017mathematics}\footnote{\url{https://github.com/springer-math/Mathematics-of-Epidemics-on-Networks}}. We extended the simulation with multiple containment strategies described in this paper. We use the R programming language for batch running the simulations by invoking the Python modules and for analyzing and plotting the simulation results.

\noindent{\bf Code availability} The source code is available at \url{https://github.com/qykong/testing-strategies}. The simulation code and all parameters used for related works can be found in the repository.

\clearpage

\clearpage


\clearpage

\clearpage
\newpage
\bibliographystyle{ACM-Reference-Format}
\bibliography{bibliography}

\newpage

\clearpage
\onecolumn
\setcounter{figure}{0}
\begin{center}
  \huge\textbf{Supplementary Figures}\\
\end{center}
\begin{figure}[H]
    \begin{subfigure}{\textwidth}
        \centering
        \includegraphics[width=0.8\textwidth,page=1]{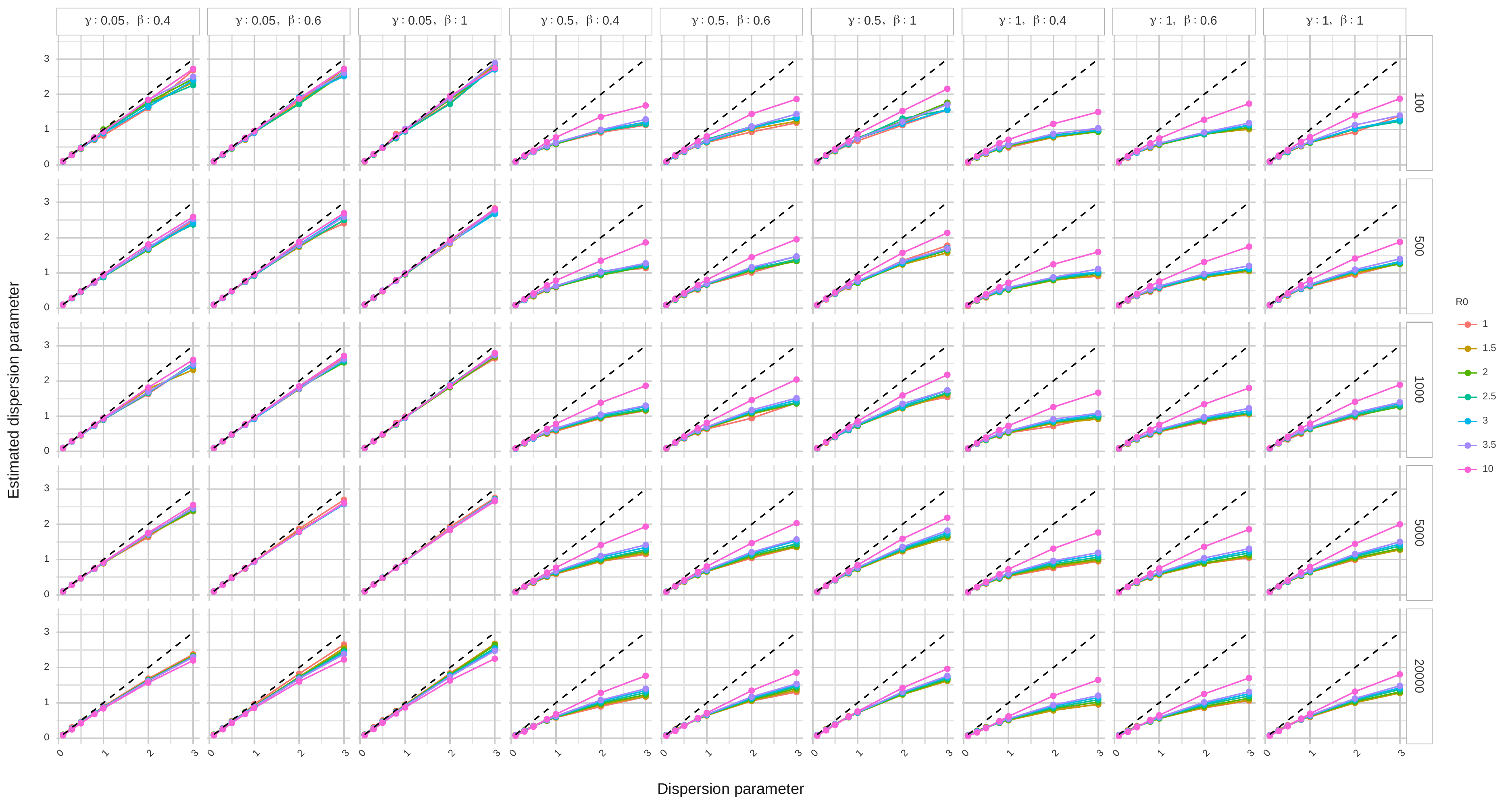}
        \caption{}
        \label{fig:network_properties_1}
    \end{subfigure}
    \begin{subfigure}{\textwidth}
        \centering
        \includegraphics[width=0.8\textwidth,page=1]{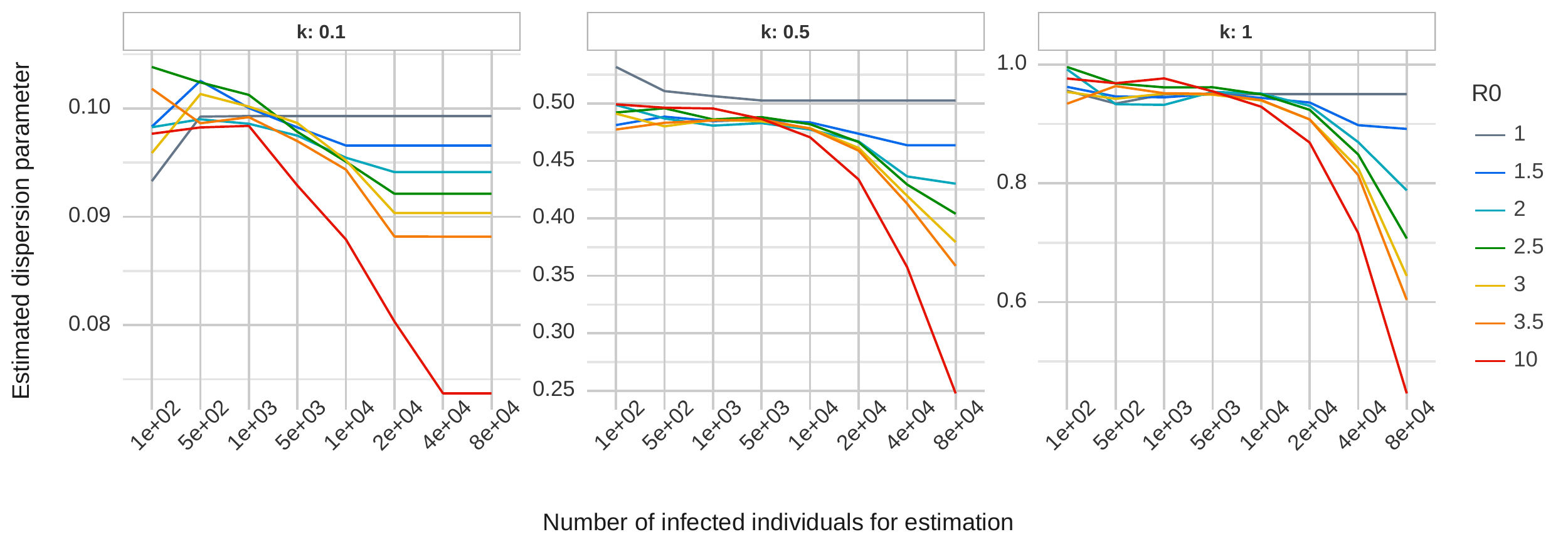}
        \caption{}
        \label{fig:network_properties_3}
    \end{subfigure}
    \caption{Chosen dispersion parameters and their empirical values. (a) evaluates the effect of computing empirical dispersion parameters over different numbers of first infected nodes. (b) shows the evolution of estimated dispersion parameters over the course of the whole epidemics for $\beta=1$ and $\gamma=0.05$.}
	\label{fig:network_properties}
\end{figure}

\begin{abstract}\end{abstract}

\begin{figure*}
    \begin{subfigure}{\textwidth}
        \centering
        \includegraphics[width=0.7\textwidth,page=1]{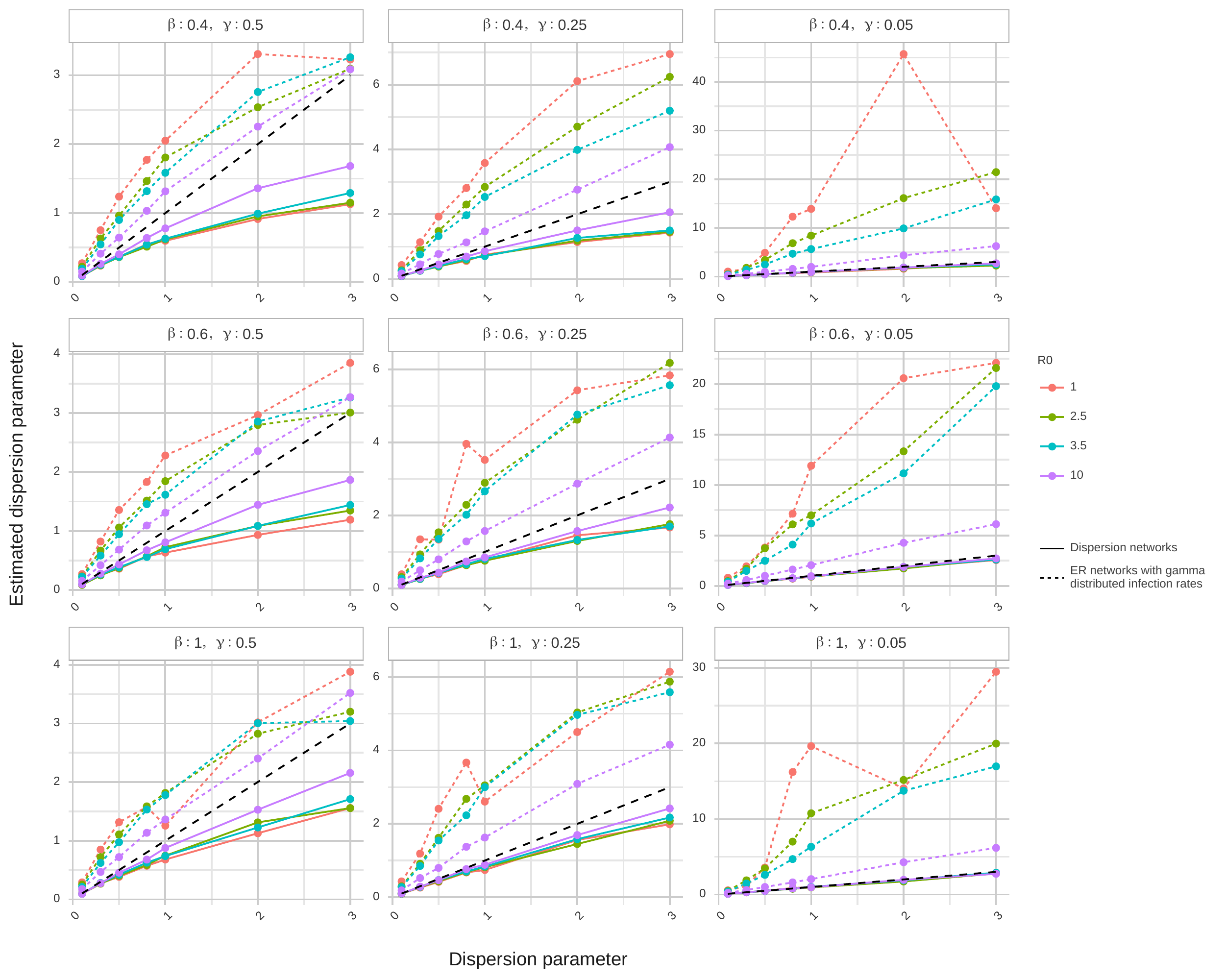}
    \end{subfigure}
    \caption{Comparing empirical dispersion parameters obtained from superspreading networks (solid colored lines) and ER networks with gamma distributed infection rates (dotted colored lines). The dotted black lines indicate the true dispersion parameters.}
    \label{fig:network_properties_2}
\end{figure*}

\begin{figure*}
    \begin{subfigure}{\textwidth}
        \includegraphics[width=1\textwidth,page=1]{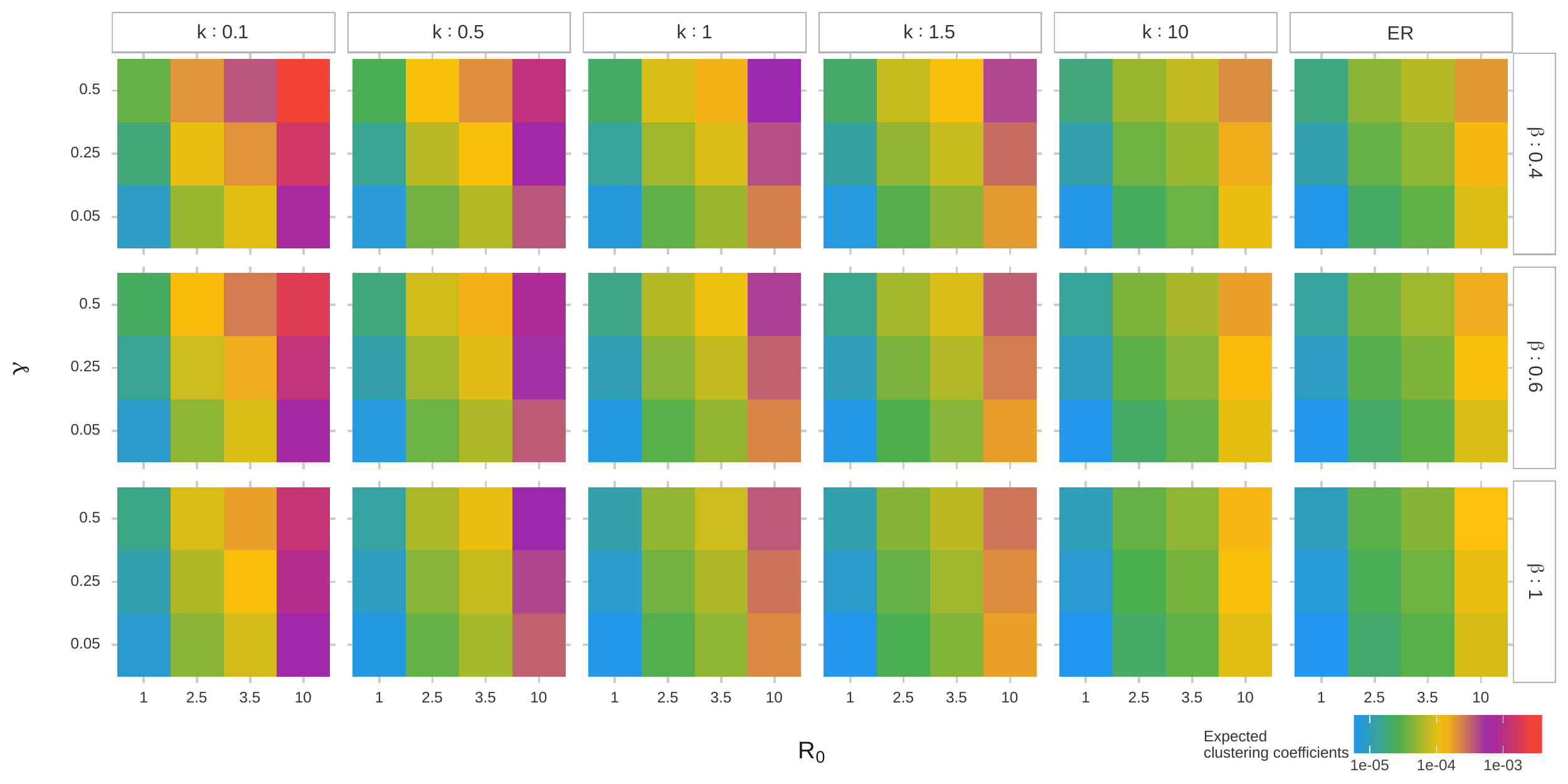}
        \caption{}
    \end{subfigure}
    \begin{subfigure}{\textwidth}
        \includegraphics[width=1\textwidth,page=1]{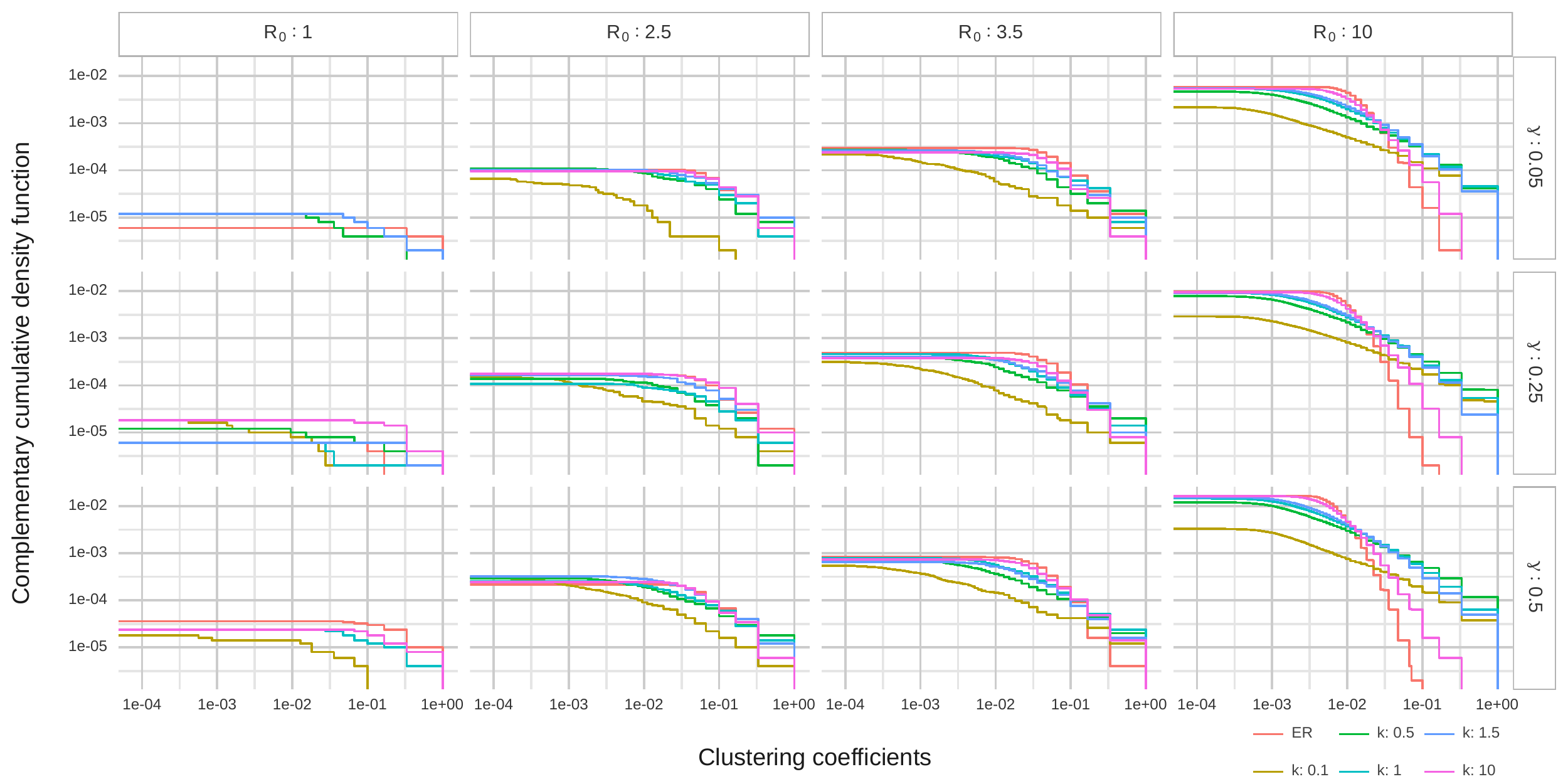}
        \caption{}
    \end{subfigure}
    \caption{Clustering coefficients given different parameters. (a) Expected clustering coefficients given different parameters when $N=100000$; (b) cumulative density plots of clustering coefficients given different parameters, $\beta=1$ }
    \label{fig:network_properties_clustering_coefficients}
\end{figure*}

\begin{figure*}[!tbp]
    \centering
    \makebox[0.5\linewidth][c]{
    \begin{subfigure}{0.9\textwidth}
    \includegraphics[width=\textwidth,page=3]{images/SEIR_positive_correlation_change_varying_parameters_selected_main_figure_beta_0.6}
    \caption{}
    \label{fig:complete_parameter_test_main_1_seir}
    \end{subfigure}
    }
    \makebox[0.5\linewidth][c]{
    \begin{subfigure}{0.9\textwidth}
    \includegraphics[width=\textwidth,page=2]{images/SEIR_positive_correlation_change_varying_parameters_selected_main_figure_beta_0.6}
    \caption{}
    \label{fig:complete_parameter_test_main_2_seir}
    \end{subfigure}
    }
    \makebox[0.5\linewidth][c]{
    \begin{subfigure}{0.9\textwidth}
    \includegraphics[width=\textwidth,page=1]{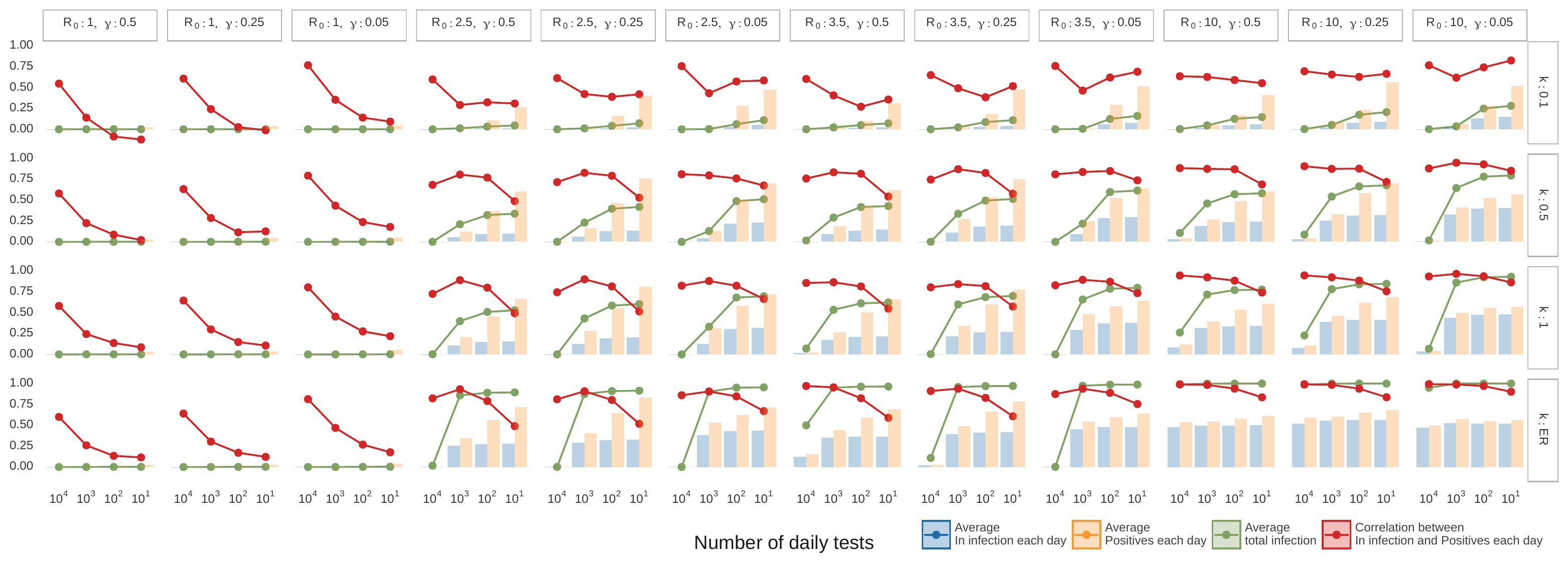}
    \caption{}
    \label{fig:complete_parameter_test_main_3_seir}
    \end{subfigure}
    }
    \caption{
        Correlation plots between daily infections and positive rates for SEIR models with different parameters, $\beta=0.6$, $\kappa=0.2$ and $P_H=0.05$. (a) shows correlation plots for random testing, (b) shows correlation plots for forward contact tracing, and (c) shows correlation plots for backward contact tracing. Red lines represent correlation values and green lines indicate expected final total infections. Bars represent expected daily infection ratios and daily average positive rates. Note that bars are scaled by the maximum values.
    }
	\label{fig:complete_parameter_test_main_seir}
\end{figure*}

\begin{figure*}[!tbp]
    \centering
    \makebox[0.5\linewidth][c]{
    \begin{subfigure}{0.9\textwidth}
    \includegraphics[width=\textwidth,page=1]{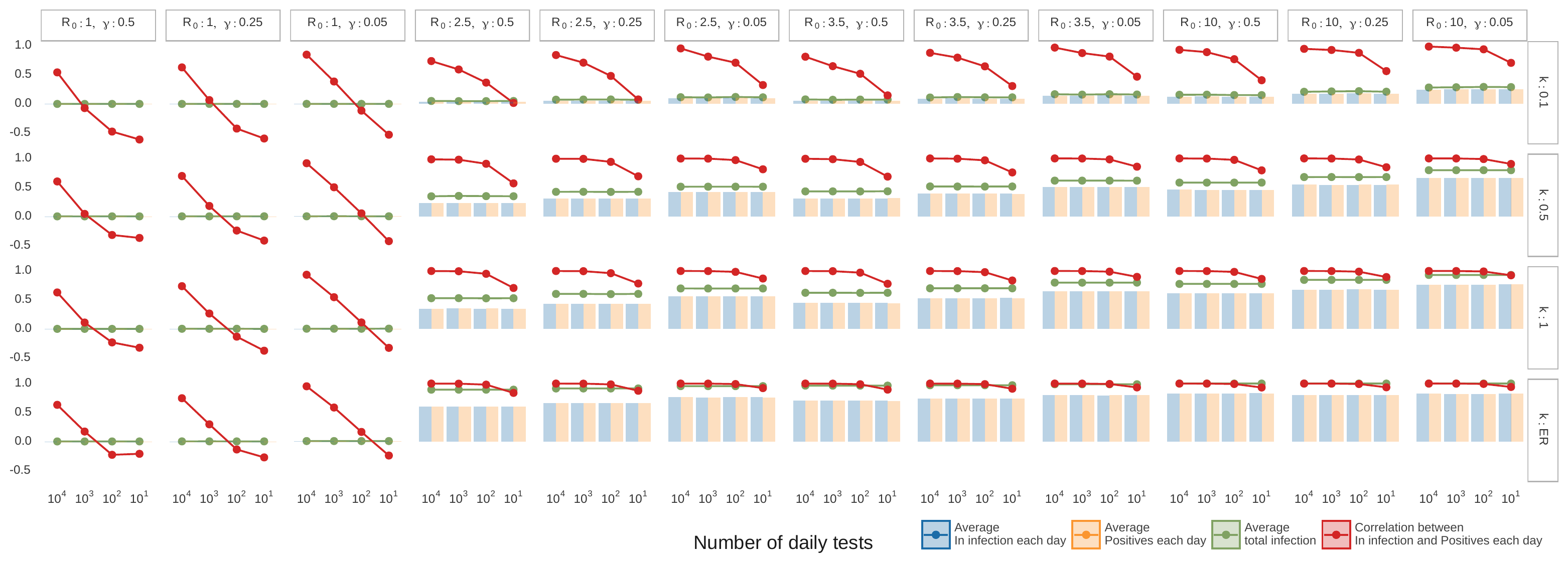}
    \caption{}
    \label{fig:complete_parameter_test_main_1_pq0}
    \end{subfigure}
    }
    \makebox[0.5\linewidth][c]{
        \begin{subfigure}{0.9\textwidth}
        \includegraphics[width=\textwidth,page=3]{images/SEIR_positive_correlation_change_varying_parameters_selected_pQ_0_2_beta_0.6}
        \caption{}
        \label{fig:complete_parameter_test_main_1_pq0_seir}
        \end{subfigure}
    }
    \caption{
        Correlation plots between daily infections and positives of (a) SIR and (b) SEIR when $P_q=0$. Note that $\beta=0.6$, $\kappa=0.2$ and $P_H=0.05$, and both plots show for results under random testing. Red lines represent correlation values and green lines indicate expected final total infections. Bars represent expected daily infection ratios and daily average positive rates. Note that bars are scaled by the maximum values.
    }
	\label{fig:complete_parameter_test_main_pq0}
\end{figure*}

\begin{figure*}[!tbp]
    \centering
    \makebox[1\linewidth][c]{
    \begin{subfigure}{0.6\textwidth}
    \includegraphics[width=\textwidth,page=1]{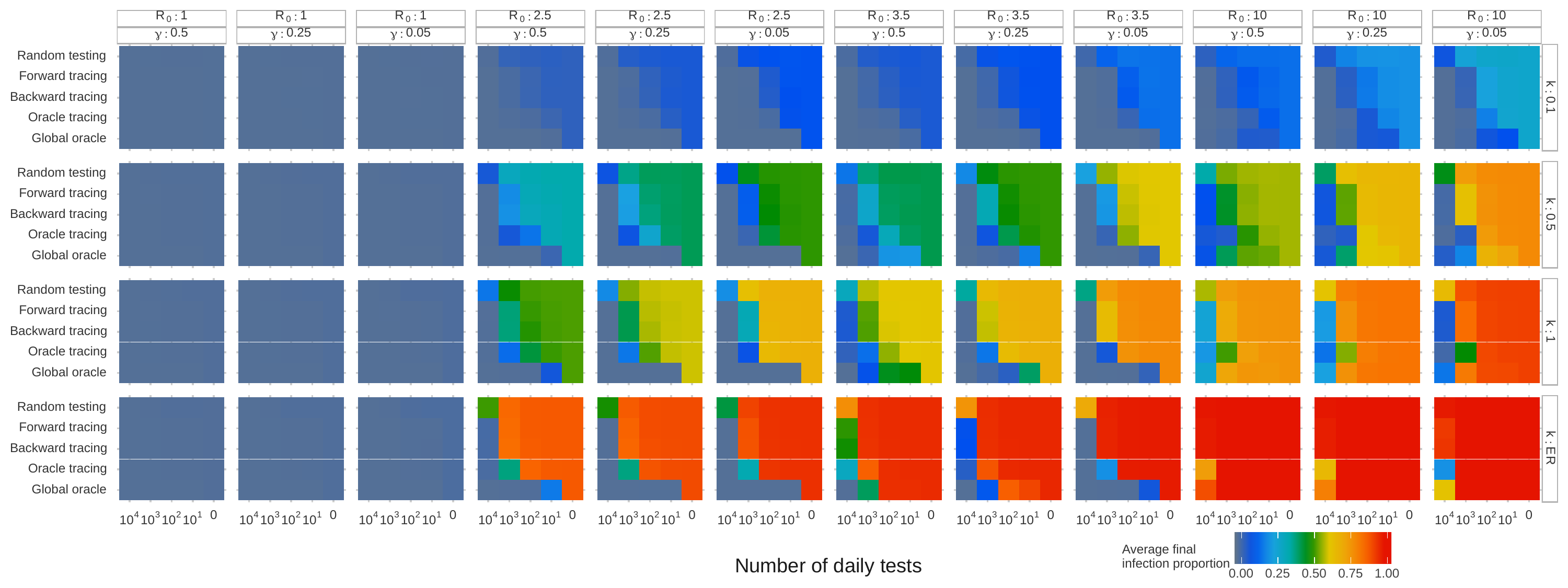}
    \caption{}
    \label{fig:complete_parameter_test_main_total_infection_1_seir}
    \end{subfigure}
    }
    \makebox[1\linewidth][c]{
    \begin{subfigure}{0.6\textwidth}
    \includegraphics[width=\textwidth,page=3]{images/SEIR_final_infection_and_days2end_main_figure_beta_0.6}
    \caption{}
    \label{fig:complete_parameter_test_main_total_infection_2_seir}
    \end{subfigure}
    }
    \makebox[1\linewidth][c]{
    \begin{subfigure}{0.6\textwidth}
    \includegraphics[width=\textwidth,page=2]{images/SEIR_final_infection_and_days2end_main_figure_beta_0.6}
    \caption{}
    \label{fig:complete_parameter_test_main_total_infection_3_seir}
    \end{subfigure}
    }
    \caption{
        Evaluation of epidemic simulations over a set of parameters for SEIR model with $\beta=0.6$, $\kappa=0.2$ and $P_H=0.05$. (a) shows average final infections, (b) shows average final infections in top 5 communities and (c) shows average days to the last infections.
    }
	\label{fig:complete_parameter_test_main_total_infection_seir}
\end{figure*}

\begin{figure*}[!tbp]
    \centering
    \makebox[1\linewidth][c]{
    \begin{subfigure}{0.6\textwidth}
    \includegraphics[width=\textwidth,page=1]{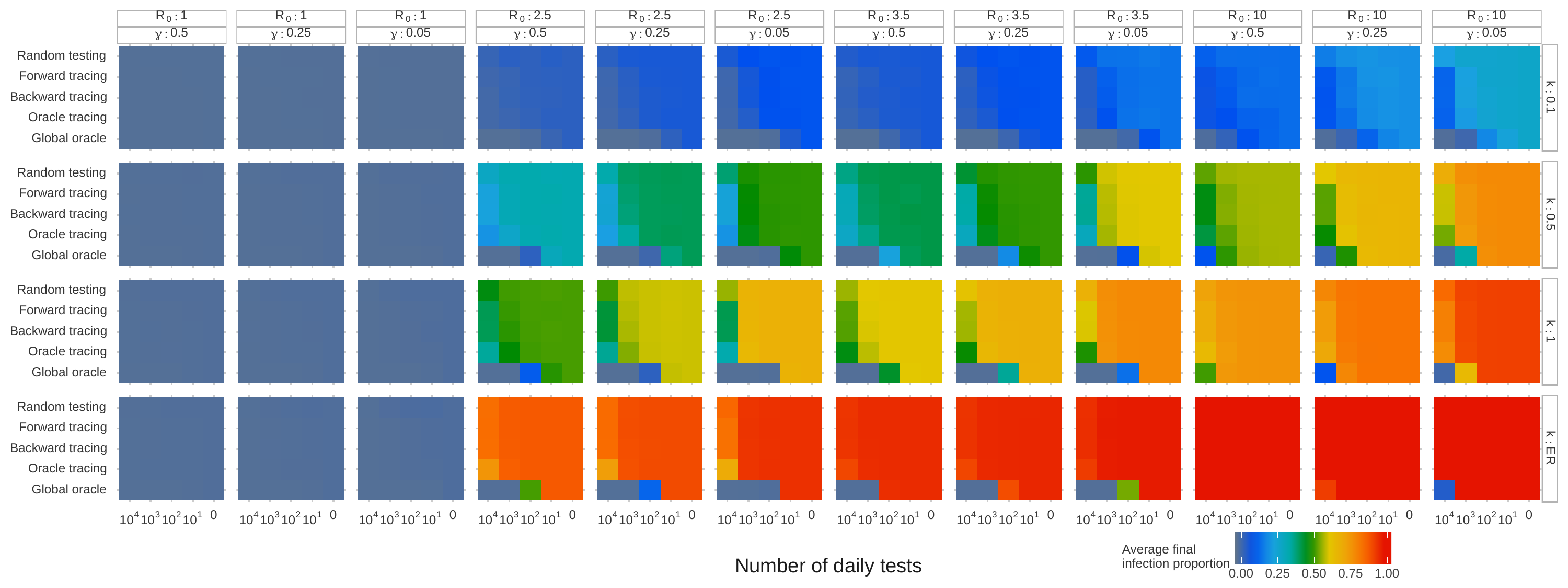}
    \caption{}
    \label{fig:complete_parameter_test_main_total_infection_1_rt100_seir}
    \end{subfigure}
    }
    \makebox[1\linewidth][c]{
    \begin{subfigure}{0.6\textwidth}
    \includegraphics[width=\textwidth,page=3]{images/SIR_final_infection_and_days2end_RT_100_tds2_beta_0.6}
    \caption{}
    \label{fig:complete_parameter_test_main_total_infection_2_rt100_seir}
    \end{subfigure}
    }
    \makebox[1\linewidth][c]{
    \begin{subfigure}{0.6\textwidth}
    \includegraphics[width=\textwidth,page=2]{images/SIR_final_infection_and_days2end_RT_100_tds2_beta_0.6}
    \caption{}
    \label{fig:complete_parameter_test_main_total_infection_3_rt100_seir}
    \end{subfigure}
    }
    \caption{
        Evaluation of epidemic simulations over a set of parameters for SIR model with $\beta=0.6$ and $P_H=0.05$ for the mixed strategy. (a) shows average final infections, (b) shows average final infections in top 5 communities and (c) shows average days to the last infections.
    }
	\label{fig:complete_parameter_test_main_total_infection_rt100_seir}
\end{figure*}

\begin{figure*}[!tbp]
    \centering
    \begin{subfigure}{0.52\textwidth}
    \includegraphics[width=\textwidth,page=1]{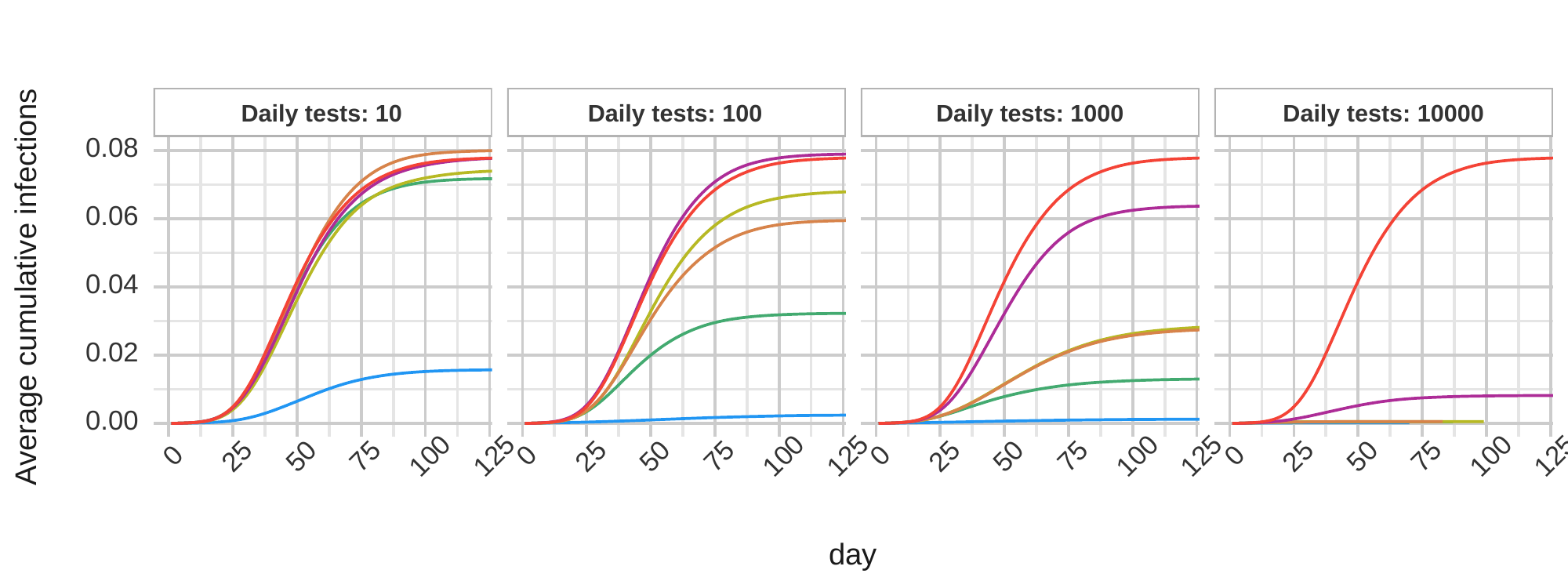}
    \caption{COVID-19}
    \label{fig:known_diseases_cum_1}
    \end{subfigure}
    \begin{subfigure}{0.52\textwidth}
    \includegraphics[width=\textwidth,page=2]{images/known_diseases_cum_infections_tmp_RT_100_2_nbinomial}
    \caption{Ebola}
    \label{fig:known_diseases_cum_2}
    \end{subfigure}
        \begin{subfigure}{0.52\textwidth}
    \includegraphics[width=\textwidth,page=3]{images/known_diseases_cum_infections_tmp_RT_100_2_nbinomial}
    \caption{SARS}
    \label{fig:known_diseases_cum_3}
    \end{subfigure}
        \begin{subfigure}{0.52\textwidth}
    \includegraphics[width=\textwidth,page=4]{images/known_diseases_cum_infections_tmp_RT_100_2_nbinomial}
    \caption{H1N1}
    \label{fig:known_diseases_cum_4}
    \end{subfigure}
        \begin{subfigure}{0.52\textwidth}
    \includegraphics[width=\textwidth,page=5]{images/known_diseases_cum_infections_tmp_RT_100_2_nbinomial}
    \caption{Measles}
    \label{fig:known_diseases_cum_5}
    \end{subfigure}
    \caption{
        The average daily cumulative infected populations of five known diseases by varying number of daily tests, under different intervention operations.
    }
	\label{fig:known_diseases_cum}
\end{figure*}

\begin{figure*}[!tbp]
    \centering
    \begin{subfigure}{0.52\textwidth}
    \includegraphics[width=\textwidth,page=1]{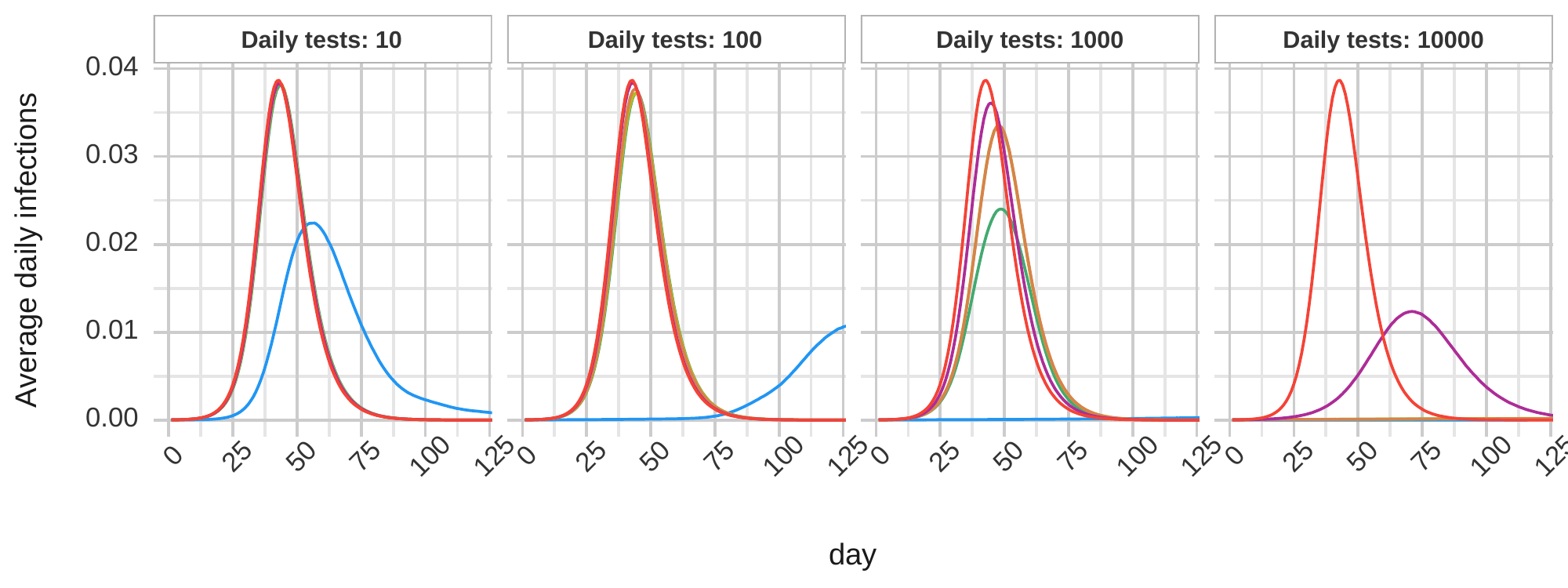}
    \caption{COVID-19}
    \label{fig:known_diseases_poisson_1}
    \end{subfigure}
    \begin{subfigure}{0.52\textwidth}
    \includegraphics[width=\textwidth,page=2]{images/known_diseases_infections_tmp_RT_100_2_poisson}
    \caption{Ebola}
    \label{fig:known_diseases_poisson_2}
    \end{subfigure}
        \begin{subfigure}{0.52\textwidth}
    \includegraphics[width=\textwidth,page=3]{images/known_diseases_infections_tmp_RT_100_2_poisson}
    \caption{SARS}
    \label{fig:known_diseases_poisson_3}
    \end{subfigure}
        \begin{subfigure}{0.52\textwidth}
    \includegraphics[width=\textwidth,page=4]{images/known_diseases_infections_tmp_RT_100_2_poisson}
    \caption{H1N1}
    \label{fig:known_diseases_poisson_4}
    \end{subfigure}
        \begin{subfigure}{0.52\textwidth}
    \includegraphics[width=\textwidth,page=5]{images/known_diseases_infections_tmp_RT_100_2_poisson}
    \caption{Measles}
    \label{fig:known_diseases_poisson_5}
    \end{subfigure}
    \caption{
        The average daily infection ratios of five known diseases by varying number of daily tests, under different intervention strategies, over ER networks.
    }
	\label{fig:known_diseases_poisson}
\end{figure*}

\begin{figure*}[!tbp]
    \centering
    \begin{subfigure}{0.52\textwidth}
    \includegraphics[width=\textwidth,page=1]{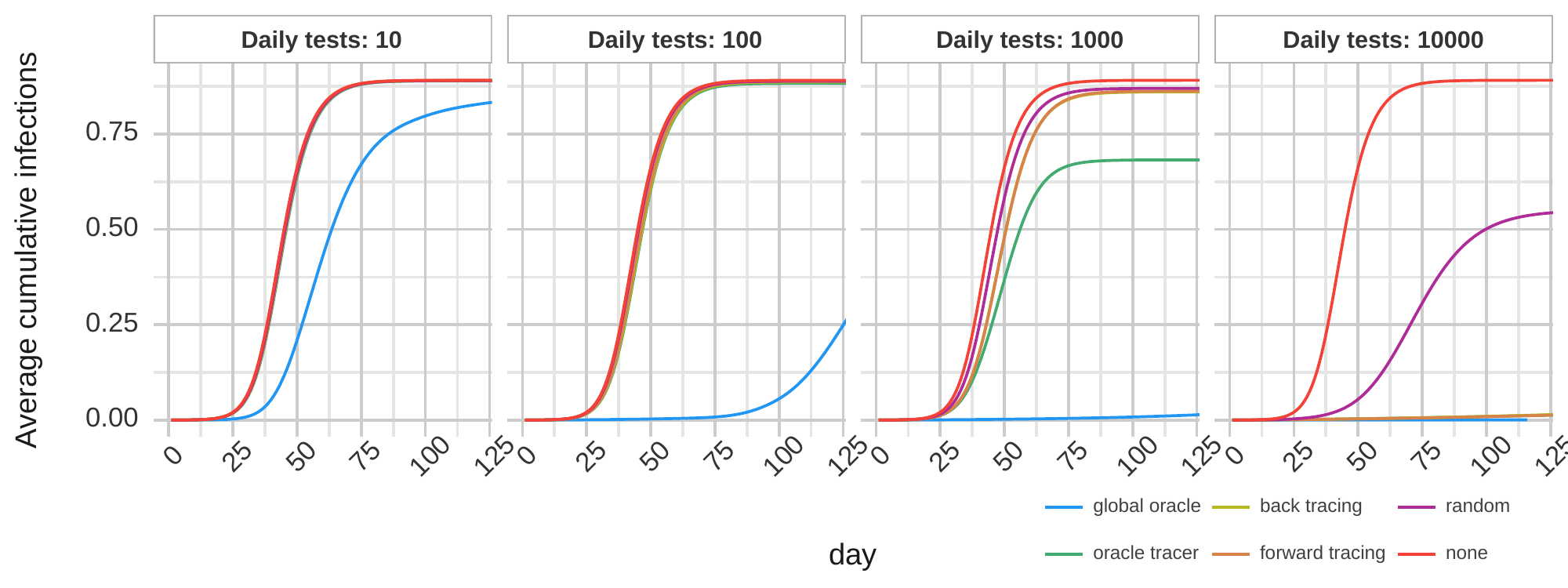}
    \caption{COVID-19}
    \label{fig:known_diseases_cum_poisson_1}
    \end{subfigure}
    \begin{subfigure}{0.52\textwidth}
    \includegraphics[width=\textwidth,page=2]{images/known_diseases_cum_infections_tmp_RT_100_2_poisson}
    \caption{Ebola}
    \label{fig:known_diseases_cum_poisson_2}
    \end{subfigure}
        \begin{subfigure}{0.52\textwidth}
    \includegraphics[width=\textwidth,page=3]{images/known_diseases_cum_infections_tmp_RT_100_2_poisson}
    \caption{SARS}
    \label{fig:known_diseases_cum_poisson_3}
    \end{subfigure}
        \begin{subfigure}{0.52\textwidth}
    \includegraphics[width=\textwidth,page=4]{images/known_diseases_cum_infections_tmp_RT_100_2_poisson}
    \caption{H1N1}
    \label{fig:known_diseases_cum_poisson_4}
    \end{subfigure}
        \begin{subfigure}{0.52\textwidth}
    \includegraphics[width=\textwidth,page=5]{images/known_diseases_cum_infections_tmp_RT_100_2_poisson}
    \caption{Measles}
    \label{fig:known_diseases_cum_poisson_5}
    \end{subfigure}
    \caption{
        The average daily cumulative infected populations of five known diseases by varying number of daily tests, under different intervention operations, over ER networks.
    }
	\label{fig:known_diseases_cum_poisson}
\end{figure*}

\begin{figure*}[!tbp]
    \centering
    \begin{subfigure}{0.62\textwidth}
    \includegraphics[width=\textwidth,page=1]{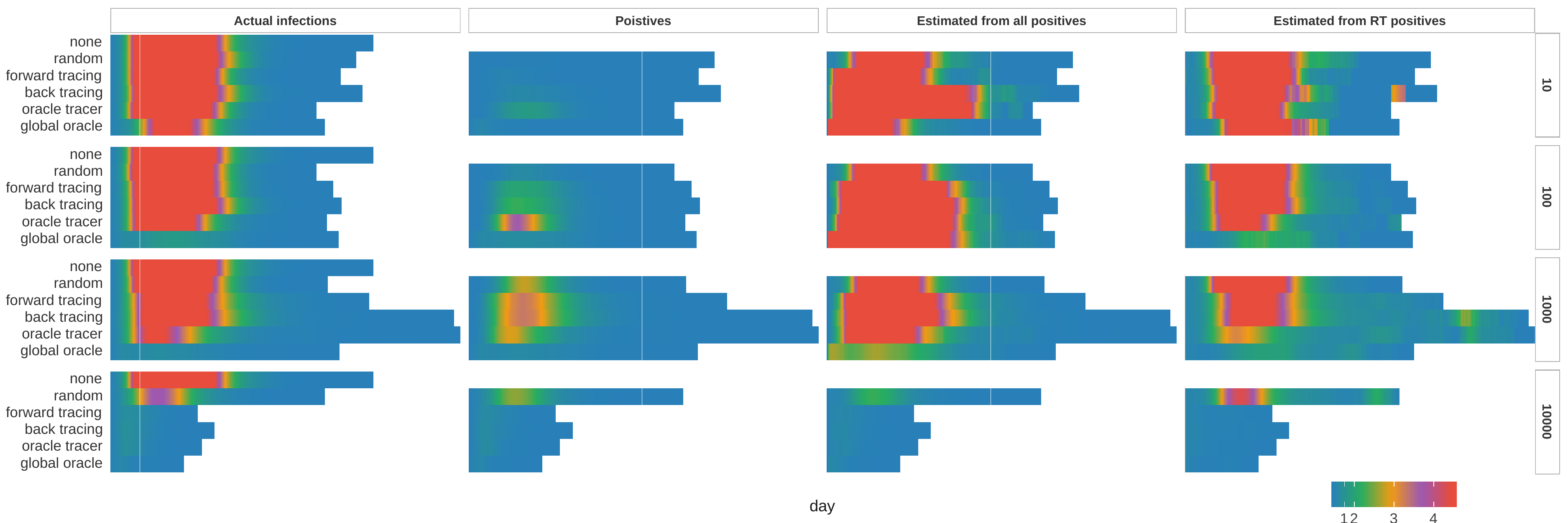}
    \caption{COVID-19}
    \label{fig:known_diseases_1_risk}
    \end{subfigure}
    \begin{subfigure}{0.62\textwidth}
    \includegraphics[width=\textwidth,page=2]{images/known_diseases_daily_alert_levels_tmp_RT_100_2_nbinomial.pdf}
    \caption{Ebola}
    \label{fig:known_diseases_2_risk}
    \end{subfigure}
        \begin{subfigure}{0.62\textwidth}
    \includegraphics[width=\textwidth,page=3]{images/known_diseases_daily_alert_levels_tmp_RT_100_2_nbinomial.pdf}
    \caption{SARS}
    \label{fig:known_diseases_3_risk}
    \end{subfigure}
        \begin{subfigure}{0.62\textwidth}
    \includegraphics[width=\textwidth,page=4]{images/known_diseases_daily_alert_levels_tmp_RT_100_2_nbinomial.pdf}
    \caption{H1N1}
    \label{fig:known_diseases_4_risk}
    \end{subfigure}
        \begin{subfigure}{0.62\textwidth}
    \includegraphics[width=\textwidth,page=5]{images/known_diseases_daily_alert_levels_tmp_RT_100_2_nbinomial.pdf}
    \caption{Measles}
    \label{fig:known_diseases_5_risk}
    \end{subfigure}
    \caption{
        The average daily threat levels of five known diseases by varying number of daily tests, under different intervention strategies.
    }
	\label{fig:known_diseases_risk}
\end{figure*}

\begin{figure*}
    \begin{subfigure}{\textwidth}
        \includegraphics[width=1\textwidth,page=1]{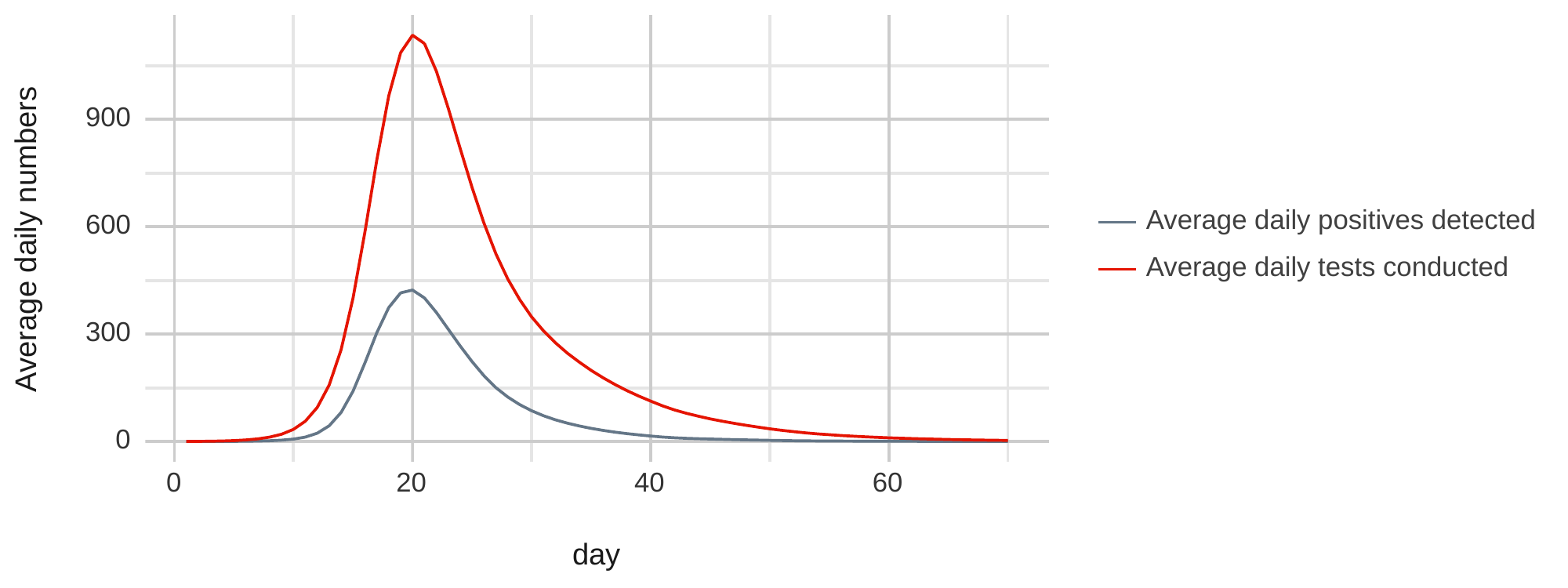}
        \caption{}
        \label{fig:aleta_model_1}
    \end{subfigure}
    \begin{subfigure}{\textwidth}
        \includegraphics[width=1\textwidth,page=2]{images/aleta_model_init_infection_1}
        \caption{}
        \label{fig:aleta_model_2}
    \end{subfigure}
    \caption{Simulations of COVID-19 on the network provided by~\citet{Aleta2020} for $I_0 = 1$. (a) shows the average number of tests used and daily positives detected when $50\%$ of symptomatic infections are discovered. (b) shows the average final infected population proportions of the scenario in (a) and, for a scenario in which only hospitalized ($2.1\%$ of symptomatic infections) are discovered (Our setup).
    Following~\citep{Aleta2020}, only $40\%$ of the contacts are discovered during contact tracing. The forward tracing strategy is considered in both figures, backward contact tracing shows identical results.}
    \label{fig:aleta_model}
\end{figure*}
\clearpage

\begin{center}
  \huge\textbf{Supplementary Tables}\\
\end{center}

\begin{table}[H]
    \caption{Parameter used for experiments}
    \label{tab:exhaustive}
    \begin{tabular}{@{}ll@{}}
    \toprule
    Parameters        & Values                    \\ 
    \midrule 
    $N$ & \textbf{100,000}, $1,000,000$ \\
    $I_0$ & \textbf{10}, 50, 100 \\
    $\beta$           & $0.2$, $0.4$, \textbf{0.6}, $1$        \\
    $\gamma$          & \textbf{0.05}, \textbf{0.25}, \textbf{0.5}, $1$       \\
    $\kappa$          & \textbf{0.2}     \\
    $R_0$             & \textbf{1}, $1.5$, $2$, \textbf{2.5}, $3$, \textbf{3.5}, \textbf{10}  \\
    $k$               & \textbf{0.1}, \textbf{0.5}, \textbf{1}, $1.5$
    \\

    $P_H$ & $0$, \textbf{0.05}, $0.1$, $0.3$   \\
    Number of daily tests & \textbf{0}, \textbf{10}, \textbf{100}, \textbf{1,000}, \textbf{10,000} \\ \bottomrule
    \end{tabular}
\end{table}
\begin{table}[H]
    \caption{Epidemic parameters of some known diseases}
    \label{tab:known_diseases}
    \begin{tabular}{llllllll}
    \toprule
        & Model & $R_0$ & $\beta$ & $\gamma$ & $\kappa$ & $H$ & $k$ \\ \midrule
    COVID-19~\citepAP{Aleta2020,endo2020estimating} & SEIR & $2.5$ & $1$ & $0.4$ & $0.2$ & $0.008372$ & $0.1$ \\
    SARS~\citepAP{chowell2003sars,Lloyd2005} & SEIR & $1.2$ & $0.15$ & $0.125$ & $0.1$ & $0.333$ & $0.16$ \\
    H1N1~\citepAP{furushima2017estimation,koliou2009epidemiological,dorigatti2012new} & SIR & $1.33$ & $0.19$ & $0.143$ & / & $0.294$ & $8.092$ \\
    Ebola~\citepAP{lekone2006statistical,althaus2015ebola} &  SEIR & $1.4$ & $0.2$ & $0.143$ & $0.2$ &  $0$ & $0.18$ \\
    Measles~\citepAP{stone2000theoretical,antona2013measles,nishiura2017assessing} & SIR & $18$ & $4.932$ & $0.274$ & / & $0.079$ & $0.32$ \\ \bottomrule
    \end{tabular}
\end{table}
\end{document}